\documentclass[prx,aps,twocolumn,superscriptaddress,longbibliography]{revtex4-1}%
\usepackage{graphicx}
\usepackage{dcolumn}
\usepackage{bm}
\usepackage{amssymb}
\usepackage{amsmath}
\usepackage{amsfonts}
\DeclareMathOperator{\trace}{Tr}
\providecommand{\U}[1]{\protect\rule{.1in}{.1in}}
\usepackage{hyperref}
\hypersetup{
colorlinks,
citecolor=blue,
filecolor=blue,
linkcolor=blue,
urlcolor=blue
}
\newcommand* {\frack}[2]{{\textstyle\frac{#1}{#2}}}

\begin{document}
%\begin{CJK*}{}{}
\title{A generalized Stoner criterion and versatile spin ordering in two-dimensional spin-orbit coupled electron systems}
\author{Weizhe Edward Liu}
\affiliation{School of Physics and Australian Research Council Centre of Excellence in Low-Energy Electronics Technologies, UNSW Node, The University of New South Wales, Sydney 2052, Australia}
\author{Stefano Chesi}
\email{stefano.chesi@csrc.ac.cn}
\affiliation{Beijing Computational Science Research Center, Beijing 100193, China}
\author{David Webb}
\affiliation{School of Physics, University of New South Wales, Sydney, NSW 2052, Australia}
\affiliation{ARC Center of Excellence for Climate System Science and Climate
Change Research Center, University of New South Wales, Sydney, NSW 2052, Australia}
\author{U.~Z\"ulicke}
\affiliation{School of Chemical and Physical Sciences and MacDiarmid Institute for Advanced Materials
and Nanotechnology, Victoria University of Wellington, Wellington 6012, New Zealand}
\author{R.~Winkler}
\affiliation{Department of Physics, Northern Illinois University, DeKalb, Illinois 60115, USA}
\author{Robert Joynt}
\affiliation{Physics Department, University of Wisconsin--Madison, Madison, Wisconsin 53706, USA}
\author{Dimitrie Culcer}
\email{d.culcer@unsw.edu.au}
\affiliation{School of Physics and Australian Research Council Centre of Excellence in Low-Energy Electronics Technologies, UNSW Node, The University of New South Wales, Sydney 2052, Australia}
\begin{abstract}
Spin-orbit coupling is a single-particle phenomenon known to generate
topological order, and electron-electron interactions cause ordered many-body phases to exist.
The rich interplay of these two mechanisms is present in a broad range
of materials, and has been the subject of considerable ongoing research and controversy. Here we demonstrate that interacting
two-dimensional electron systems with strong spin-orbit coupling exhibit a variety of time reversal symmetry breaking
phases with unconventional spin alignment. We first prove that a Stoner-type  criterion can be formulated
for the spin polarization response to an \textit{electric field}, which predicts that the spin polarization susceptibility
diverges at a certain value of the electron-electron interaction strength. The divergence indicates the
possibility of unconventional ferromagnetic phases even in the absence of any applied electric or magnetic field.
This leads us, in the second part of this work, to study interacting
Rashba spin-orbit coupled semiconductors \textit{in equilibrium} in the Hartree-Fock approximation as a generic
minimal model. Using classical Monte-Carlo simulations
we construct the complete phase diagram of the system as a function
of density and spin-orbit coupling strength. It includes both an out-of-plane spin polarized phase and
in-plane spin-polarized phases with shifted Fermi surfaces and rich spin textures, reminiscent of the Pomeranchuk instability,
as well as two different Fermi-liquid phases having one and two Fermi surfaces,
respectively, which are separated by a Lifshitz transition.
We discuss possibilities for experimental observation and useful application of these novel phases,
especially in the context of electric-field-controlled macroscopic spin polarizations.
\end{abstract}
\date{\today}
\maketitle
%\end{CJK*}

% Send paper to Tamborenea, Erez Berg, Giovanni, Raikh.

% Make clear from beginning diff between Bloch Stoner and Rashba Stoner.
% Shuffling around occupation numbers: forgot about n_k.
% Is there a \sqrt{2} added to \tilde{alpha}? In Eq. 37. The transition between FL1 and FL2 occurs at \tilde{alpha} = 1, this means \tilde{alpha} missing \sqrt{2}.

\section{Introduction}

Spin-orbit coupling manifests itself in a great variety of spin textures in
solids \cite{Winkler03, Chapman11nature, Datta90apl, Zutic04rmp, Bernevig06prl, Ghosh04prl, Shen14prl, PhysRevLett.116.246801, PhysRevB.86.081306, PhysRevLett.93.046602, PhysRevB.76.245322},
many of which are associated with topological effects and states of
matter \cite{Hasan_TI_RMP10, Qi_TI_RMP_10, Moore-nature, Koenig_HgTe_QSHE_Science07, 
Yu_TI_QuantAHE_Science10, Chang_QAHE_exper_Science2013, Oh153, 
PhysRevB.81.125332, PhysRevB.82.155457, Culcer_TI_Int_PRB11}.
Electron-electron interactions \cite{Vignale.05, PhysRevB.15.2819}, on the other hand, lead to ordered
states, including a large number of phases characterized by spin ordering \cite{Ceperley_nature, 
PhysRevB.18.3126, PhysRevLett.79.463, PhysRevLett.82.5317, PhysRevLett.88.256601, 
PhysRevLett.95.230403, PhysRevLett.90.136601, PhysRevB.67.073304, PhysRevB.78.155304, 
0295-5075-77-3-37003}.
It is therefore natural to expect systems that have strong electron-electron
interactions as well as strong spin-orbit coupling to exhibit a multitude of
exotic, unconventional states of matter. In light of this, the fascinating interplay of spin-orbit coupling and electron-electron interactions has
received considerable attention in recent years, in materials ranging from
topological insulators to conventional semiconductors \cite{PhysRevLett.108.117201, PhysRevB.60.4826, PhysRevB.72.195319, PhysRevB.85.075321, PhysRevA.90.043614, PhysRevB.88.075115, PhysRevB.88.125309, PhysRevB.84.235117, PhysRevB.83.235308, PhysRevB.83.235309, Yu2016}.

Besides the fundamental scientific importance of the subject, current interest in it is fuelled by its inherent
great potential for technological applications \cite{Sinova_SHE_RMP15}. 
Present-day information technology
is based on semiconducting and spin-based devices to store and process
information. Combining magnetic and semiconducting properties may lead to
faster and more efficient operation with minimal power consumption. This
merger was attempted in the past using ferromagnetic semiconductors, but the
achievement of larger critical temperatures $T_{c}$ required sizable
concentrations of magnetic impurities, leading to seemingly intractable
difficulties such as very low mobilities and phase separation \cite{RevModPhys.78.809, PhysRevLett.97.136603, PhysRevLett.98.146602, 
PhysRevB.69.174412, PhysRevB.75.214420, 
PhysRevB.76.054424, PhysRevB.79.155208, PhysRevB.86.094406, Stagraczyński201679, PhysRevB.95.054422}. One aim of the present work is to perform a conceptual study to determine whether, in a series of selected model
systems, spin-orbit coupling can provide an avenue to combine magnetic and
semiconducting properties without resorting to magnetic doping. In this context the larger question is whether spin-orbit coupling can be harnessed to generate and
preserve spin polarizations \textit{in equilibrium}, which in the long run could foster the development of long-sought spintronics applications.
In this work we answer the above questions in two steps.

We first study the spin polarization induced by an electric field in a Rashba spin-orbit coupled system in the presence of electron-electron interactions. The current-induced spin polarization is also known as the Edelstein effect \cite{EDELSTEIN1990233, Aronov_89, 
PhysRevLett.93.176601, Yu_APL04, PhysRevB.82.195316, PhysRevLett.109.246604}, and a spin \textit{polarization} is understood as an average over the individual spin \textit{orientations} of the electrons in all occupied states. Our transport formalism captures spin-orbit coupling, disorder and driving fields on the same footing, while treating electron-electron interactions in the Hartree-Fock approximation. We find that a generalized Stoner criterion can be formulated
for the electrically-induced spin polarization, predicting that at a certain
interaction strength the spin density response to an external electric field
diverges. Beyond this point in parameter space the spin polarization is
sensitive to an infinitesimal electric field and as a result the system is
expected to develop a spontaneous in-plane spin polarization in equilibrium,
in a manner reminiscent of the Bloch transition to a ferromagnetic phase.

The emergence of a dramatic interaction enhancement for the electrically induced spin polarization
motivates a focus on the equilibrium state in the second step.
We investigate the ground state of the same interacting system in the absence of an electric field
in the Hartree-Fock approximation. Although this configuration has been
studied in the past \cite{gfg_proc05, Juri2005, Chesi07prb, Juri_PRB08, Zhang16pra, 
Ruhman_PRB14, Barnes_SR14, Kim16prb},
previous works either focused entirely on phases in which the location of
the Fermi surface remains fixed \cite{gfg_proc05,Juri2005, Chesi07prb,Juri_PRB08,Zhang16pra}
and which are consequently similar in nature
to the well-known Stoner ferromagnetism \cite{Doniach.74, Marder.10},
or did not consider the competition of exotic ferromagnetic phases with more conventional ones
\cite{Ruhman_PRB14, Barnes_SR14,Kim16prb}.
Our present investigation is driven by the expectation
that, since the appearance of an electrically-induced spin polarization is
qualitatively different from the generation of net spin densities via the
Zeeman effect, the equilibrium phase that is expected to emerge when the
corresponding response function diverges will be qualitatively different from ordinary
Stoner ferromagnetism. In the second step of this work therefore we present
mean-field analytical and numerical calculations of the interacting ground
state of a Rashba spin-orbit coupled semiconductor. Our equilibrium formalism utilizes classical Monte-Carlo
simulations to solve the Hartree-Fock equations. Where possible we compare the
Monte-Carlo results to analytical approximations, obtaining excellent agreement.

\begin{figure}[t]
\begin{center}
\includegraphics[width=1.03\columnwidth]{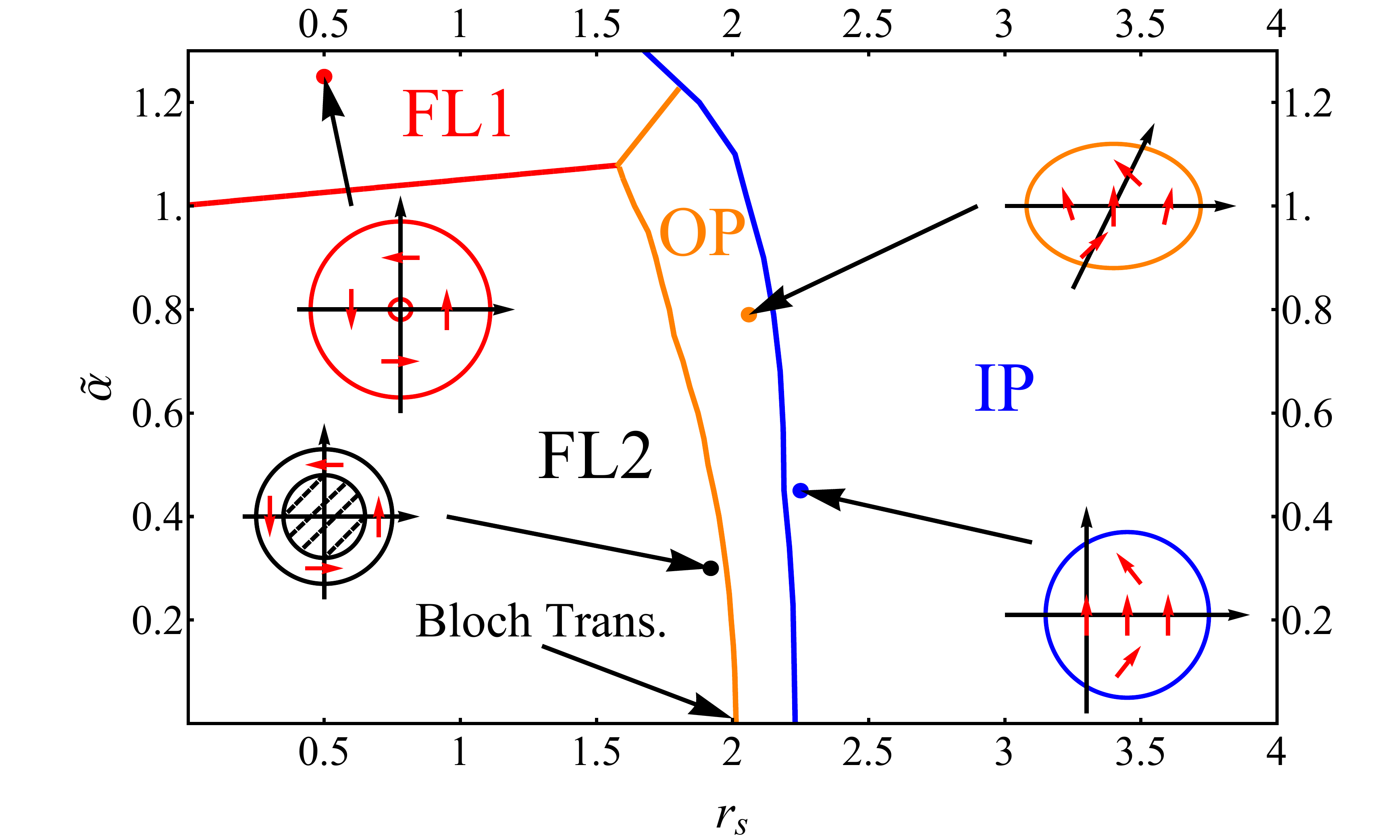}
\end{center}
\caption{Phase diagram of 2D electron liquid with Rashba
spin-orbit coupling obtained by solving the Hartree-Fock equations using a Monte Carlo method.
Here $\tilde\alpha$ and $r_s$ are dimensionless measures for the strength of Rashba
spin-orbit coupling and electron-electron interactions, respectively: $\tilde\alpha$ corresponds to
the ratio of Fermi wave length and spin-precession length, and $r_s$ is the Wigner-Seitz radius
of the 2D electron system. The distinguishing ground-state features for each individual phase
are indicated schematically. For the paramagnetic Fermi-liquid phase FL1 (FL2), there is no
net spin polarization, and the ground state is a Fermi sea formed from one (both) spin
subband(s). In contrast, the OP phase is characterized by a centered Fermi surface and an
out-of-plane magnetization. The IP phase is the most unconventional, exhibiting an in-plane
magnetization associated with a shifted Fermi sea.
}%
\label{MonteCarlo_final}%
\end{figure}

A central result of our work is a detailed map of the entire Hartree-Fock phase diagram for the interacting spin-orbit-coupled system shown in Fig.~\ref{MonteCarlo_final}. It reveals a rich diversity of phases involving out-of-plane and in-plane spin polarized phases even at relatively small values of the Wigner-Seitz radius $r_{s}$, as well as a Lifshitz transition \cite{Lifshitz1} between spin-unpolarized Fermi liquid phases with one and two Fermi surfaces respectively. As expected from the non-equilibrium calculation performed in the first step the spin-orbit-induced in-plane spin-polarized state is characterized by a Fermi surface shifted away from the origin. We further show that at high values of the spin-orbit coupling and
electron-electron interaction strengths the Fermi surface is both shifted away from the origin and distorted, in a manner that recalls the Pomeranchuk instability \cite{Pomeranchuk1, PhysRevLett.85.5162, PhysRevB.63.153103, PhysRevLett.102.116404, PhysRevLett.98.126407, Quintanilla20081279, PhysRevB.72.035114, PhysRevB.74.115126}.

The spin-polarized phases that we identify differ qualitatively from the customary Stoner ferromagnetism. Specifically, Stoner ferromagnetism involves an abrupt transition from an unpolarized to a fully spin-polarized phase (i.e. occupation numbers at most equal to unity) without any intermediate partially-polarized phases. The spin polarization may point in any direction, and there is no spin texture in reciprocal space. In contrast, the spin-polarized phases that we describe in this work can be either fully spin-polarized or partially polarized (involving occupation numbers of 0, 1 or 2). Their spin polarization can be out-of-plane, in-plane or \textit{tilted} in reciprocal space, and are generally accompanied by a complex spin texture in reciprocal space, which is directly related to the spin texture of the Rashba spin-orbit effective field in the absence of electron-electron interactions. To begin with, an out-of-plane spin polarization emerges at relatively small values of $r_s$ as a result of the fact that electrons save exchange energy by lining up their spins parallel to each other. This can only occur in a direction perpendicular to the plane so as to avoid the effect of the Rashba interaction. However, a weak angular structure of the spin polarization in reciprocal space reflects the presence of the Rashba spin-orbit coupling \cite{gfg_proc05, Juri_PRB08}. At larger values of $r_s$ the in-plane spin polarized phase emerges, which involves the creation of a spontaneous net spin-orbit effective field and is accompanied by a sizable shift in the Fermi surface. The spin texture of this phase becomes exceedingly complex as $r_s$ and the spin-orbit coupling strength increase. 

The calculations presented in this work are based on the paradigmatic~\cite{Bihlmayer15newjphys} Rashba Hamiltonian as a minimal model of a two-dimensional (2D) spin-orbit coupled semiconductor. This choice is motivated in part by the expectation that the proven ability~\cite{TuneRashba1, TuneRashba2, TuneRashba3} to tune independently both the 2D-electron density and the structural inversion asymmetry that gives rise to the Rashba spin-orbit coupling using external electric fields will enable experimental access to,
and novel technological exploitation of, the identified unconventional phases. However, we expect our qualitative findings to hold quite generally in 2D semiconductors with strong spin-orbit interactions \cite{PhysRevLett.93.226602, PhysRevB.70.155308, PhysRevB.78.235417, PhysRevB.84.235411, PhysRevB.87.085408, PhysRevB.87.245121, Siegert_NP07}.

The emergence of a net effective magnetic field is familiar from the customary description of electrically-induced spin polarizations, which we recall occurs in gyrotropic materials \cite{Ivchenko_78, Vorobev_79} in a non-equilibrium diffusive regime \cite{EDELSTEIN1990233, Aronov_89}. Stoner ferromagnetism, on the other hand, has no such symmetry restrictions. Hence we expect in-plane spin-polarized phases with shifted Fermi surfaces to emerge in systems displaying electrically-induced spin polarizations. Whereas most 2D spin-orbit models do give rise to electrically-induced spin polarizations, we stress that the in-plane polarized phases with shifted Fermi surfaces we reveal in this work are not a universal feature of spin-orbit coupled systems. In addition to the requirement of gyrotropic symmetry, it is also necessary for the system to have two Fermi surfaces in the non-interacting state. In analogy with Stoner ferromagnetism, one of these can be regarded as the \textit{minority} spin sub-band and the other as the \textit{majority} spin sub-band. Electron-electron interactions enhance the contribution to the spin polarization stemming from the majority sub-band and reduce that of the minority sub-band. However, a system in which spin-orbit coupling is dominant, such as a topological insulator, has a single Fermi surface, which corresponds to the minority spin sub-band. Hence we do not expect exotic in-plane spin-polarized phases in topological insulators. 

The effects we predict are readily observable in experiments. The net spin-orbit effective field singles out a spatial direction. When a small in-plane external magnetic field is applied we expect an anisotropy in the resistance as the magnetic field is rotated in the plane of the 2DEG. The system will display an anomalous Hall effect as well, which we expect to be rather complex in nature, involving a net effective magnetic field at each point on the Fermi surface. The calculation of this effect will need to be performed separately. 

The organization of this paper is as follows. In Sec.~\ref{sec:transport}, we develop our density matrix formalism for the interacting Rashba spin-orbit coupling system under an applied electric field. In Sec.~\ref{Interacting_SP}, we analytically study the current-induced spin polarization in the interacting Rashba system and determine the exact expression for the spin polarization. In Sec.~\ref{sec:equi}, we introduce the numerical methods used in studying the ground state of the interacting Rashba system.
In Sec.~\ref{sec:NumRes}, we numerically determine the phase diagram of the interacting Rashba system,
in which an in-plane spin polarized ground state emerges under certain circumstances.
In Sec.~\ref{sec:small_alpha}, we give a complete analytical treatment of the small spin-orbit coupling limit.
The results are discussed in Sec.~\ref{sec:disc}. We end with a summary and outlook.

\section{Density Matrix Formalism}

\label{sec:transport}

In this and next section, we will perform the analytical calculation on the Coulomb interaction effect
on the current-induced spin polarization in a 2D Rashba spin-orbit coupled electron system,
determining the spin susceptibility to an electric field in the presence of electron-electron interactions.
We only focus on the density matrix formalism in this section.

\subsection{Hamiltonian}

\label{Hamiltonian}

The many-body Hamiltonian is
\begin{equation}\label{Hmany}
H = \sum_{\bm{k} \bm{k}' s s'} \bigg[
\langle s | H^\mathrm{sp}_{\bm{k}\bm{k}'} |s'\rangle \, c^{\dagger}_{\bm{k} s} c_{\bm{k}' s'}
\bigg] + V^\mathrm{ee}\, ,
\end{equation}
where $c_{\bm{k} s}$ is the annihilation operator for a single electron with wavevector ${\bm k}$ and spin index $s=\pm$, and $c_{\bm{k} s}^\dagger$ is the corresponding creation operator. The single-particle Hamiltonian is \cite{PhysRevB.78.235417}
\begin{equation}
H^\mathrm{sp}_{\bm{k}\bm{k}'} = H_{0{\bm k}} \delta_{{\bm k}{\bm k}'} + U_{{\bm k}{\bm k}'} + H_{E,{\bm k}{\bm k}'}\, ,
\end{equation}
where the three contributions, discussed in the following, are the band Hamiltonian $(H_{0})$, the coupling to the electric field $(H_\mathrm{E})$ and a
disorder potential accounting for short-range scattering $(U)$. The Coulomb interaction term $V^\mathrm{ee}$  will be discussed below. 

In the presence of Rashba spin-orbit coupling the band Hamiltonian takes the following form in the crystal-momentum representation
\begin{equation}
\label{RashbaHamiltonian}H_{0\bm{k}} = \frac{\hbar^{2} k^{2}}{2 m} + \alpha\,
\bm{\sigma}\cdot(\bm{k} \times \hat{\bm{z}})\, ,
\end{equation}
where $m$ is the band effective mass, $\alpha$ is the Rashba coefficient,
$\bm{\sigma} = (\sigma_{x}, \sigma_{y}, \sigma_{z})$ is the vector of Pauli matrices,
$\bm{k} = (k_{x}, k_{y})$ is the in-plane wavevector, and
$\hat{\bm{z}}$ is a unit vector along the $\bm{z}$ direction. It is possible to include the Dresselhaus spin-orbit coupling \cite{Yu2016}, but it is not the focus of this paper.
There are two spin-split bands: the upper band with energy $\varepsilon_{\bm{k}+}$ and the lower band with energy $\varepsilon_{\bm{k}-}$,
where $\varepsilon_{\bm{k} \pm} = \hbar^{2} k^{2} / 2 m \pm \alpha k$
and $k = \vert \bm{k} \vert$. For the lower band, there is a ring of energy minima $\varepsilon_\mathrm{min} = - \alpha^{2} m / 2 \hbar^{2} $ at $k = k_c \equiv \alpha m / \hbar^{2}$.

There are two possible Fermi sea configurations, depending on whether the upper
band is occupied or not. At low density, only the lower band is occupied and
the Fermi sea consists of a single annulus with inner radius
$k_\mathrm{F,i}$ and outer radius $k_\mathrm{F,o}:$
\begin{equation}
\label{oneFermi surfacekfs}
k_\mathrm{F,i} = k_{c} - \frac{\pi n \hbar^{2}}{m \alpha}\, ;\quad 
k_\mathrm{F,o} = k_{c} + \frac{\pi n \hbar^{2}}{m \alpha}\, ,
\end{equation}
where $n$ is the electron density. From Eq.~(\ref{oneFermi surfacekfs}), we see that a critical carrier density
$n_\mathrm{c} = k_c^2 / \pi = m^{2} \alpha^{2} / \pi\hbar^{4}$ may be defined corresponding to the point at which $k_\mathrm{F,i} \ge 0$.
When the density exceeds $n_\mathrm{c}$, the upper band is also occupied and
the Fermi sea is composed of two circular Fermi disks with opposite spin
alignments. The radii of the upper and lower disks are, respectively,
\begin{equation}\label{kfp}
k_{\mathrm{F}+} \!= \! \sqrt{2 \pi n \!- k_{c}^{2} } - k_{c}\, ; \quad
k_{\mathrm{F}-} \! = \! \sqrt{2 \pi n \! - k_{c}^{2} } + k_{c}\, .
\end{equation}

We consider a uniform electric field applied in the plane of the sample. Working in the \textit{Pauli basis} of spin eigenstates of the matrix $\sigma_z$, the potential $e{\bm E}\cdot\bm{r}$ describing the coupling to the electric field $\bm{E} = (E_{x},E_{y})$, with $\bm{r}$ the position operator and $e$ the elementary charge, is simply represented by
\begin{equation}\label{ElectricField_Exp}
H_{\mathrm{E}\bm{k}} =  i e \bm{E} \cdot \frac{\partial}{\partial \bm{k}}\, .
\end{equation}

The disorder potential in real space is conventionally written as
\begin{equation}
U (\bm{r}) = \sum_{I} \, \mathcal{U} (\bm{r} - \bm{R}_{I})\, ,
\end{equation}
where summation over $I$ indicates the inclusion of all impurities and $\bm{R}_{I}$ is the impurity coordinate.
The configuration average of the short-range disorder potential $U$ in the reciprocal space is
\begin{equation}
\overline{\rule{0pt}{2ex}\langle \bm{k} \vert U \vert \bm{k}' \rangle \langle \bm{k}' \vert U \vert \bm{k} \rangle}
= n_{I} \vert \mathcal{U}_{\bm{k}\bm{k}'} \vert^{2} / A\, ,
\end{equation}
where $n_{I}$ is the impurity density and $A$ is the total area.

The Coulomb interaction $V^\mathrm{ee}$
takes the standard form
\begin{equation}\label{Vee_exp}
V^\mathrm{ee} = \frac{1}{2A} \sum_{\bm{k} \bm{k}' s s'} \sum_{\bm{q} \ne 0}
V_{\bm{q}} \, c^{\dagger}_{\bm{k} + \bm{q},s}
c^{\dagger}_{\bm{k}' - \bm{q},s'} c_{\bm{k}'s'} c_{\bm{k} s}\, .
\end{equation}
In 2D, the screened Coulomb potential
matrix element for momentum transfer $\bm{q} = \bm{k} - \bm{k}'$ is
\begin{equation}\label{Vee_begin}
V_{\bm{q}} = \frac{e^{2}}
{2\varepsilon_{r} \varepsilon_{0} (k_\mathrm{TF}+ |\bm{q}|)}\, ,
\end{equation}
where $k_\mathrm{TF}$ is the Thomas-Fermi wavenumber and $\varepsilon_{r}$ is the static dielectric constant. Finally, the Wigner-Seitz radius $r_{s}$ is introduced by
\begin{equation}
\label{eq:Wigner-Seitz}
r_{s} = \frac{m e^{2}}{4 \pi \varepsilon_{r} \varepsilon_{0} \hbar^{2} \sqrt{\pi n}}\, ,
\end{equation}
which represents the relative strength of the electron-electron interactions to
the average kinetic energy. We note that, although strictly speaking the Wigner-Seitz radius is poorly defined in multiband systems, here we use the definition of $r_s$ for $\alpha = 0$ purely as a convenient dimensionless parameter to quantify the strength of the electron-electron interactions.

\subsection{Kinetic equation}

\label{GeneralFormalism}

We follow the density matrix formalism for the kinetic equation of the interacting systems
\cite{Vasko, PhysRevB.84.235411, PhysRevB.87.085408}.
The quantum Liouville equation for the many-body density matrix $F$ is
\begin{equation}\label{MB_LE}
\frac{\mathrm{d} F}{\mathrm{d} t} + \frac{i}{\hbar} [H,F] = 0\, ,
\end{equation}
where $H$ is the many-body Hamiltonian (\ref{Hmany}). The one-particle reduced density matrix is
\begin{equation}\label{OP_DM_Exp}
\rho_{{\bm k}{\bm k}'}^{ss'} = \trace (c_{{\bm k}'s'}^{\dagger} c_{{\bm k}s} F)\, ,
\end{equation}
where $\trace$ is the trace over all variables including momenta and spins.
By employing Wick's theorem we obtain an effective single-particle kinetic equation \cite{PhysRevB.84.235411}
\begin{equation}\label{sp_keq}
\frac{\mathrm{d} f_{\bm{k}}}{\mathrm{d} t} + \frac{i}{\hbar} [H_{0\bm{k}},f_{\bm{k}}]
+ J (f_{\bm{k}}) = \frac{e \bm{E}}{\hbar} \cdot \frac{\partial f_{\bm{k}}}{\partial \bm{k}}
+ \frac{i}{\hbar} [\mathcal{B}_{\bm{k}},f_{\bm{k}}],
\end{equation}
where $f_{\bm{k}}$ is the $\bm{k}$-diagonal part of the single-particle density matrix $\rho$,
$J (f_{\bm{k}})$ is the Born scattering term due to the impurity scattering potential $U$,
and $\mathcal{B}_{\bm{k}}$ is the Hartree-Fock mean-field Coulomb interaction.
Note that $f_{\bm{k}}$ is a $2\times2$ matrix since we do not write the spin indices explicitly. The interplay of electron-electron interactions and disorder involve only the ${\bm k}$-diagonal part of $\rho$. The contribution of its $\bm{k}$-off-diagonal part is associated with Altshuler-Aronov corrections
\cite{PhysRevLett.112.146601, PhysRevB.60.5818}, which is not the focus of this article.
Similarly, weak localization and antilocalization corrections to the semiclassical limit are not included in our study, which allows us to write $J (f_{\bm{k}})$ in the first-order Born approximation as follows
\cite{PhysRevB.78.235417}
\begin{equation}\label{scatterm}
\arraycolsep 0.1em
\begin{array}[b]{r>{\displaystyle}l} \displaystyle
J(f_{\bm{k}}) = \frac{n_{I}}{\hbar^{2}} \lim_{\delta \to 0} &
\int \frac{\mathrm{d} \bm{k}'}{(2\pi)^{2}} \; |\mathcal{U}_{\bm{k}\bm{k}'}|^{2}
\int_{0}^{\infty} \mathrm{d} t \, \mathrm{e}^{-\delta t} \\[3ex]
\times \big\{ & \mathrm{e}^{- i H_{0\bm{k}'} t / \hbar} 
 (f_{\bm{k}} - f_{\bm{k}'}) \mathrm{e}^{ i H_{0\bm{k}} t /\hbar} \\[1ex]
 & {} + \mathrm{e}^{- i H_{0\bm{k}} t /\hbar} (f_{\bm{k}} - f_{\bm{k}'})
\mathrm{e}^{i H_{0\bm{k}'} t / \hbar} \big\}.
\end{array}
\end{equation}
The electric-field driving term $D^{E}_{\bm{k}} \equiv (e \bm{E}/\hbar) \cdot (\partial f_{\bm{k}} / \partial \bm{k})$
is nonzero when the term~(\ref{ElectricField_Exp}) is included in $H$. The mean-field (exchange) Coulomb interaction $\mathcal{B}_{\bm{k}}$ can be written as
\begin{equation}\label{Bk_exp}
\mathcal{B}_{\bm{k}} (f_{\bm{k}}) = \int \frac{\mathrm{d} \bm{k}' }{(2\pi)^{2}} \; V_{\bm{k} - \bm{k}'} f_{\bm{k}'},
\end{equation}
and then $\trace [(-\mathcal{B}_{\bm{k}}) f_{\bm{k}} ]$ gives the expression of the exchange energy
by using Eq.~(\ref{Vee_begin}) [which corresponds to the exchange energy appearing below in Eq.~(\ref{exchangestart})].

\section{Interacting Spin-orbit Coupled Electrons in an Electric Field}

\label{Interacting_SP}

For a Rashba spin-orbit coupled system, the electrically induced spin polarization is calculated by solving the kinetic equation~(\ref{sp_keq}). The dynamics of the spin-density matrix can be derived for a general spin-orbit coupled system. We will concentrate on the zero temperature case in this section.

\subsection{General decomposition of the density matrix}

Most generally the density matrix can be decomposed as $f_{\bm k} = n_{\bm k} \openone + S_{{\bm k}\parallel} + S_{{\bm k}\perp}$. The scalar contribution $n_{\bm k}$ represents the charge density, while $S_{\bm{k} \parallel}$ is the fraction of the spin density at each ${\bm k}$ parallel to the Rashba effective field $\sigma_{\bm{k} \parallel} = \bm{\sigma}\cdot(\hat{\bm{k}} \times \hat{\bm{z}})$, and $S_{\bm{k} \perp}$ is the part perpendicular to the Rashba field. We can write $S_{\bm{k} \perp} \propto \sigma_{\bm{k} \perp}$, where 
$\sigma_{\bm{k} \perp} = \bm{\sigma} \cdot \hat{\bm{k}}$ is orthogonal to $\sigma_{\bm{k}\parallel}$ in the sense that ${\rm tr} \sigma_{\bm{k} \perp} \sigma_{\bm{k}\parallel} = 0$, with tr the spin trace.

The philosophy of our approach can be summarized as follows. We begin with the non-interacting system, so $\mathcal{B}_{\bm{k}}$ is zero. To achieve this formally one could let e.g. the relative permittivity $\varepsilon_r \rightarrow \infty$. In the absence of an external electric field there is no net spin polarization in the system, whereas when an electric field ${\bm E}$ is applied to the non-interacting system it gives rise to a finite electrically-induced spin polarization. At this point we turn on the electron-electron interaction, and in the presence of a nonzero electrically-induced spin polarization $\mathcal{B}_{\bm{k}}$ itself becomes nonzero. This nonzero field is inserted into Eq.~(\ref{sp_keq}), which is then solved to yield an additional contribution to the spin polarization, which in turn gives rise to a new contribution to $\mathcal{B}_{\bm{k}}$, and this self-consistent process is iterated in search of a closed-form solution. 

It is important to note that (i) $\mathcal{B}_{\bm{k}}$ enters the kinetic equation through the commutator $[\mathcal{B}_{\bm{k}}, f_{\bm k}]$ on the right-hand side, and that (ii) $\mathcal{B}_{\bm{k}} \propto {\bm E}$. Within the framework of linear electric-field response one may therefore replace $f_{\bm k}$ inside the commutator by the equilibrium density matrix, given below, which commutes with the spin-orbit Hamiltonian. This implies immediately that all terms involving the identity matrix and $\sigma_{{\bm k}\parallel}$ drop out of the commutator, and the electron-electron interaction correction only affects $S_{{\bm k}\perp}$. As a result there is \textit{no enhancement} of the charge density. Only the spin density is enhanced, that is, the difference between spin-up and spin-down.

\subsection{Noninteracting case}

Without the electron-electron interactions, $\mathcal{B}_{\bm{k}} = 0$ in Eq.~(\ref{sp_keq}).
The transport equation~(\ref{sp_keq}) is then much simplified,
and can be evaluated by using Eq.~(\ref{RashbaHamiltonian}).
In this section, we will assume $ m \alpha / \hbar^{2} k_\mathrm{F} \ll 1$
so a perturbative treatment of $\alpha$ is used to calculate the spin polarization.

The Fermi-Dirac distribution function of the Rashba system takes the general form
\begin{equation}\label{DF_full}
f_{0\bm{k}} = \frac{f_{0\bm{k}}^{+} + f_{0\bm{k}}^{-}}{2} + \bm{\sigma}\cdot(\hat{\bm{k}} \times \hat{\bm{z}})
\frac{f_{0\bm{k}}^{+} - f_{0\bm{k}}^{-}}{2}\, ,
\end{equation}
where $f_{0\bm{k}}^{\pm} = \Theta (\varepsilon_\mathrm{F} - \varepsilon_{\bm{k}\pm})$
with $\Theta (x)$ is the step function at $x = 0$
and $\hat{\bm{k}}$ is an unit vector along the $\bm{k}$ direction.
The Fermi level of the Rashba system $\varepsilon_\mathrm{F} = \hbar^{2} k_\mathrm{F}^{2} / 2 m + m\,\alpha^{2}/\hbar^{2}$,
where $k_\mathrm{F} = \sqrt{ 2 \pi n}$ is the Fermi wavenumber in the absence of spin-orbit coupling.
In the leading order in $\alpha$, Eq.~(\ref{DF_full}) becomes
\begin{equation}\label{f0k_appr}
f_{0\bm{k}} \approx \Theta \bigg(\frac{\hbar^{2}k_\mathrm{F}^{2}}{2 m} - \varepsilon_{0\bm{k}} \bigg)
- \varepsilon_{\mathrm{SO},\bm{k}}\,  \delta \bigg(\frac{\hbar^{2} k_\mathrm{F}^{2}}{2m} - \varepsilon_{0\bm{k}}\bigg),
\end{equation}
where $\varepsilon_{0\bm{k}} = \hbar^{2} k^{2} / 2m$ and
$\varepsilon_{\mathrm{SO},\bm{k}} = \alpha\, \bm{\sigma}\cdot(\bm{k} \times \hat{\bm{z}})$
is the Rashba spin-orbit term.

The applied electric field will induce a correction to the density matrix $f_{E\bm{k}}$, so the total density matrix will be written as
\begin{equation}\label{densitycomp}
f_{\bm{k}} = f_{0\bm{k}} + f_{E\bm{k}}\, ,
\end{equation}
with corresponding decompositions for $n_{\bm k}$ and $S_{\bm k}$. The scattering term (\ref{scatterm}) can be decomposed as $J(f_{\bm{k}}) = J(n_{\bm k}) + J(S_{\bm{k}\parallel}) + J(f_{\bm{k}\perp})$, while $J(f_{0\bm{k}}) = 0$.
In the order of $\alpha^{0}$, we obtain, from Eq.~(\ref{scatterm}),
\begin{equation}\label{Jfpara_exp}
J(n_{\bm k} + S_{\bm{k}\parallel}) = \frac{2 m \, n_{I} u^{2}}{\hbar^{3}}
\int \frac{\mathrm{d} \theta'}{2\pi} [(n_{\bm k} + S_{\bm{k} \parallel}) - (n_{{\bm k}'} + S_{\bm{k}' \parallel})] \, ,
\end{equation}
where we write $\mathcal{U}_{\bm{k}\bm{k}'} \equiv u$ for short-range impurities. For the model we study it is safe to write Eq.~(\ref{Jfpara_exp})
as $J (n_{\bm k} + S_{\bm{k}\parallel}) = (n_{\bm k} + S_{\bm{k}\parallel}) / \tau_{0}$ where $\tau_{0}^{-1} = m\, n_{I} u^{2} / \hbar^{3}$.

In linear response the driving term $D^E_{\bm{k}}$ that is due to the electric field can be approximated as
$D^{E}_{\bm{k}} \approx (e \bm{E} / \hbar ) \cdot (\partial f_{0\bm{k}} / \partial \bm{k})$
whose parallel component to the Rashba field is, in the leading order in $\alpha$,
\begin{equation}
D^{E}_{\bm{k}\parallel} = - \frac{e \alpha m \, \bm{E} \cdot \hat{\bm{k}}}{\hbar^{2}} \sigma_{\bm{k} \parallel}
\bigg[ \frac{\partial}{\partial k} \delta (k - k_\mathrm{F}) - \frac{1}{k} \delta (k - k_\mathrm{F})\bigg],
\end{equation}
Note that, up to linear order in $\alpha$, there is no $\sigma_{\bm{k} \perp}$ component in $D^{E}_{\bm{k}}$.
The solution of the kinetic equation (\ref{sp_keq}) is simply $S_{\bm{k}\parallel} = \tau_{0} D^{E}_{\bm{k} \parallel}$,
so the non-interacting $y$-direction spin polarization is
\begin{equation}\label{sy0_exp}
s_{y}^{(0)} = \int \frac{A \, \mathrm{d} \bm{k}}{(2\pi)^{2}} \, \trace \big(\frack{1}{2} \hbar \, \sigma_{y} S_{\bm{k} \parallel} \big)
=- \frac{e \,\alpha E_{x} A \,m \,\tau_{0}}{2 \pi \hbar^{2}}.
\end{equation}
This is the current-induced spin polarization due to the Rashba spin-orbit coupling \cite{Ivchenko_78, Vorobev_79, EDELSTEIN1990233, Aronov_89, 
PhysRevLett.93.176601, Yu_APL04, PhysRevB.82.195316, PhysRevLett.109.246604}. Note that Eq.~(\ref{sy0_exp}) also matches the expression for the 2D Dirac fermions \cite{PhysRevB.78.235417} in which the Rashba spin-orbit coupling is the dominant term in the Hamiltonian.

\subsection{Interaction enhancement of the current-induced spin polarization}

\label{InteractingSP}

In the presence of the electron-electron interaction,
we will add $\mathcal{B}_{\bm{k}}$ [see Eq.\ (\ref{Bk_exp})] to the kinetic equation~(\ref{sp_keq})
and then self-consistently solve for the density matrix.
We will keep the leading order in $\alpha$ in the spin polarization,
and observe a divergence when the electron-electron strength $r_{s}$ exceeds a critical value.

The self-consistency in solving the kinetic equation~(\ref{sp_keq}) relies on the iteration
of the density matrix solution $S_{\bm{k}\perp}$, as shown in Refs.~\cite{PhysRevB.84.235411, PhysRevB.87.085408}.
We start the first iteration by setting $f_{\bm{k}} = f_{0\bm{k}}$, which corresponds to the non-interacting results.
The resulting first-order term in $\mathcal{B}_{\bm{k}}$ takes the form
\begin{equation}
\mathcal{B}^{(1)}_{\bm{k}} = \frac{e^{3} \tau_{0} E_{x} \alpha m}{4 \pi \varepsilon_{r} \varepsilon_{0} \hbar^{3} k_\mathrm{F}}
\, \mathrm{I}_{2} (k) \sin \theta \, \sigma_{\bm{k} \perp},
\end{equation}
where we only keep the $\sigma_{\bm{k} \perp}$ component because the parallel component drops out of the commutator with $H_{0{\bm k}}$.
The definition of $\mathrm{I}_{2} (k)$ is
\begin{equation}
\arraycolsep 0.3ex
\begin{array}[b]{rl}
\displaystyle \mathrm{I}_{2} (k) = \int^{2 \pi}_{0} & \displaystyle \frac{\mathrm{d} \gamma}{2 \pi}
\frac{k_\mathrm{F} \sin^{2} \gamma}{k_\mathrm{TF} + q(k,k_\mathrm{F},\gamma)} \\[3ex]
\times & \displaystyle \bigg[ \, 2 - \frac{k_\mathrm{F} (k_\mathrm{F} - k \cos \gamma)}{ [ k_\mathrm{TF} + q (k, k_\mathrm{F},\gamma)]
q (k, k_\mathrm{F},\gamma)}\bigg],
\end{array}
\end{equation}
where $q(k,k_\mathrm{F},\gamma) = \sqrt{k^{2} + k_\mathrm{F}^{2} - 2 k k_\mathrm{F} \cos \gamma}$.
The driving term due to the electron-electron interaction becomes
\begin{subequations}
\label{Dee_1}
\begin{align}
 D^{ee, (1)}_{\bm{k}} = & \frac{i}{\hbar} [\mathcal{B}_{\bm{k}}^{(1)},f_{\bm{k}}] \approx \frac{i}{\hbar} [\mathcal{B}_{\bm{k}}^{(1)},S_{0\bm{k}}] \\
= & \frac{\tau_{0}\, \alpha^{2} e^{3} E_{x} m^{2}}{2\pi \varepsilon_{r} \varepsilon_{0}\hbar^{6} k_\mathrm{F}}
\mathrm{I}_{2} (k) \, \delta(k - k_\mathrm{F}) \sin \theta \, \sigma_{z},
\end{align}
\end{subequations}
where $S_{0\bm{k}}$ is the spin-dependent part of $f_{0\bm{k}}$ [see Eq.~(\ref{f0k_appr})], which may be displayed as
\begin{equation}
S_{0\bm{k}} = - \frac{ m \alpha}{\hbar^{2}} \bm{\sigma}\cdot(\hat{\bm{k}} \times \hat{\bm{z}}) \, \delta (k_\mathrm{F} - k)\, .
\end{equation}

With $D^{ee, (1)}_{\bm{k}}$ appearing on the right-hand side of the kinetic equation~(\ref{sp_keq}),
we obtain
\begin{equation}\label{Dee1_eq}
\frac{\mathrm{d} S_{\bm{k}\perp}^{(1)}} {\mathrm{d} t}
+ \frac{i}{\hbar} \Big[ H_{0\bm{k}}, S_{\bm{k}\perp}^{(1)} \Big] = D^{ee, (1)}_{\bm{k}}.
\end{equation}
The solution of Eq.~(\ref{Dee1_eq}) is found straightforwardly as
\begin{subequations}
\begin{align}
S_{\bm{k}\perp}^{(1)} = & \lim_{\delta \to 0^{+}} \int_{0}^{\infty} \mathrm{e}^{- i H_{0\bm{k}} t / \hbar } \,
D^{ee, (1)}_{\bm{k}} \, \mathrm{e}^{i H_{0\bm{k}} t / \hbar } \, \mathrm{e}^{- \delta t} \mathrm{d} t\, , \\
 = & - \frac{\tau_{0} \alpha e^{3} E_{x} m^{2}}{4 \pi \varepsilon_{r} \varepsilon_{0} \hbar^{5} k_\mathrm{F} k}
\mathrm{I}_{2} (k) \delta (k - k_\mathrm{F}) \sin \theta \sigma_{\bm{k} \perp}.
\end{align}
\end{subequations}
Then the first-order interaction enhancement of the spin polarization is
\begin{equation}\label{sy1_exp}
s_{y}^{(1)} = \int \frac{A \, \mathrm{d} \bm{k}}{(2\pi)^{2}}  \trace
\big( \frack{1}{2} \hbar \sigma_{y} S_{\bm{k} \perp}^{(1)} \big) 
= \frac{r_{s} \mathrm{I}_{2} (k_\mathrm{F}) }{2 \sqrt{2}} s^{(0)}_{y} \equiv \lambda_{1} s^{(0)}_{y}.
\end{equation}

The second iteration is performed as
\begin{equation}
\mathcal{B}^{(2)}_{\bm{k}} = \mathcal{B}_{\bm{k}} (S_{\bm{k}\perp}^{(1)})
= \frac{e^{3} \tau_{0} E_{x} \alpha m}{2 \pi \varepsilon_{r} \varepsilon_{0} \hbar^{3} k_\mathrm{F}}
\lambda_{1}
\, \mathrm{I}_{3} (k) \sin \theta \, \sigma_{\bm{k} \perp},
\end{equation}
where the definition of $\mathrm{I}_{3} (k)$ is
\begin{equation}
\mathrm{I}_{3} (k) = \int^{2 \pi}_{0} \! \frac{\mathrm{d} \gamma}{2 \pi}
\frac{k_\mathrm{F} \cos^{2} \gamma}{k_\mathrm{TF} + q(k,k_\mathrm{F},\gamma)}\, .
\end{equation}
For $s_{y}^{(2)}$ we obtain
\begin{equation}
s_{y}^{(2)} = r_{s} \mathrm{I}_{3} (k_\mathrm{F} )s_{y}^{(1)} / \sqrt{2} \equiv \lambda_{2} s_{y}^{(1)}.
\end{equation}
For the third iteration, we will write out $S_{\bm{k}\perp}^{(2)} = \lambda_{2} S_{\bm{k}\perp}^{(1)}$, and finally we get $s_{y}^{(3)} = \lambda_{2} s_{y}^{(2)}$. Thus, for the $n$-th $(n>0)$ iteration, we have $s_{y}^{(n)} = \lambda_{2} s_{y}^{(n-1)} = \lambda_{2}^{n-1} s_{y}^{(1)}$.

To summarize, after having considered the electron-electron interaction,
the corresponding spin density corrections, which only act parallel to the spin-orbit Hamiltonian,
are then iteratively calculated by including an effective Hamiltonian~(\ref{Bk_exp}) as the driving source.
The correction to the spin-polarization stemming from electron-electron interactions is represented by a geometric series, which can be summed exactly.
The total spin polarization of the system can be written as
\begin{equation}\label{interactSP_final}
s_{y} = s^{(0)}_{y} + \sum_{n=1}^{\infty} \lambda_{2}^{n-1} s_{y}^{(1)}
\equiv \left( 1 + \frac{\lambda_{1}}{1-\lambda_{2}} \right) s^{(0)}_{y},
\end{equation}
with $\lambda_{2} \le 1$.  From Eq.~(\ref{interactSP_final}), the spin polarization $s_{y}$ will diverge if $r_{s} = \sqrt{2} / \mathrm{I}_{3} (k_\mathrm{F})$. Thus, whereas the Edelstein effect reflects a small perturbation in response to the electric field, the divergence in this response signals a sizable enhancement. In other words, when the electron-electron interactions become sufficiently large, the spin polarization of the system will respond to any small electric field. The response function characterizing the Edelstein effect is proportional to the product of the spin-orbit constant $\alpha$ and the scattering time $\tau$, and we note that both of these drop out of the condition for the divergence of the spin polarization. If we compare this to the divergence of the Zeeman response to a magnetic field, leading to the customary Stoner criterion for ferromagnetism, it is evident that the role of the magnetic field in our setup is taken over by the electric field, while the quantity $\alpha\tau$ plays the role of the $g$-factor. In fact, one way to visualize this effect is to consider the spin-orbit coupling, the electric field and the scattering time as giving rise to a net effective magnetic field \cite{OptOrient}. It is the spin response to this magnetic field that diverges. In contrast to Stoner ferromagnetism, the spin polarization here is not free to point in any direction, but is constrained to lie in the plane because the net effective magnetic field lies in the plane. Note once more that the divergence occurs only in the spin-dependent part of the response function, not the charge part. The latter is not renormalized by electron-electron interactions. 

These results suggest that at a certain interaction strength the system becomes susceptible to infinitesimally small external electric fields. This in turn suggests that the system tends to develop a net in-plane spin polarization \textit{in the absence} of an external electric field. Moreover, since the non-equilibrium spin polarization is accompanied by a shift in the Fermi surface away from the Brillouin zone center leading to the formation of a net spin-orbit effective field, we expect the equilibrium spin-polarized phase to have a Fermi surface displaced from ${\bm k} = 0$ and a non-trivial spin texture. In other words we expect the system to develop an equilibrium phase with a nonzero in-plane spin polarization that is physically distinct from Stoner ferromagnetism and is associated with the creation of a net spin-orbit effective magnetic field, whose spin texture may be rather complex. The near-equilibrium approach we have pursued so far cannot give us further insight, and to determine the conditions for the existence of this equilibrium phase as well as its qualitative nature we need to examine the Hartree-Fock phase diagram of the interacting system \textit{in equilibrium}. 

\section{Interacting spin-orbit coupled electrons in equilibrium}

\label{sec:equi}

We have seen that a small electric field, whose effect is to shift the Fermi surface, can induce a spin polarization in an interacting spin-orbit coupled system.  This immediately suggests the possibility that interactions alone could shift the Fermi surface. The resulting state would carry no electrical current, a fact that we will demonstrate explicitly below, but would nevertheless have a net spin polarization. This is in addition to the theoretical background for the 2D electron gas with no spin-orbit coupling, which has a transition to a ferromagnetic state at low density.  Thus there are two candidates for a spin-polarized state: the out-of-plane (OP) state \cite{Juri_PRB08, gfg_proc05} and the in-plane (IP) state. The terminology refers to the spin directions relative to the plane of the 2D system. We find that these OP and IP states compete in a nontrivial way and both appear in our final phase diagram, shown in Fig.~\ref{MonteCarlo_final}.
% Guide the reader through the phenomenology introduced by SO on top of the Bloch ferromagnet. Previously magnetization can point anywhere, now only OP. 
The OP state may be thought of as a spin texture that interpolates smoothly between a purely ferromagnetic state with all spins pointing in the $+z$ direction, which minimizes the exchange energy, and the unpolarized Rashba-spin-split Fermi-sea state, which minimizes the spin-orbit energy. Given this description, we expect the OP state to become favored as the density decreases, and this is indeed seen in Fig.~\ref{MonteCarlo_final}. The way this evolution takes place in momentum space will emerge below in Fig.~\ref{OP_state}.

The IP state is actually a collection of spin textures whose precise configuration depends on spin-orbit strength and density, all of which are characterized by the spin direction lying in the plane of the 2D system.   While the OP state is in some sense a perturbative modification of the venerable Bloch ferromagnet by the spin-orbit coupling, the IP state is more exotic.
Its existence is perhaps best understood in the ideal limit of large spin-orbit coupling $k_c \gg k_F $, when the Fermi disk moves way off the center of the Brillouin zone to a location around $k=k_c$. While maintaining a nearly circular occupation has a small non-interacting energy cost [compared to the annulus of Eq.~(\ref{oneFermi surfacekfs})], the spin-orbit field is almost constant on the displaced Fermi disk and does not compete with the exchange field. Thus, ferromagnetism is favored by both exchange and the spin-orbit coupling, which is the basic mechanism that drives the IP phase.  Hence we expect low density and large spin-orbit coupling to drive the IP phase, and this is reflected in the large area of the phase diagram (see Fig.~\ref{MonteCarlo_final}) that the IP phase occupies.

Our only assumption on the magnetic ordering of the ground state is that each ${\bm k}$-state corresponds to a definite spin direction.  This does exclude some types of spin-density-wave (SDW) states~\cite{Marinescu_PRB15}. Although SDW states can have a lower exchange energy than the paramagnetic (PM) state~\cite{Overhauser_PRL60}, SDW phases are usually disfavored by correlation effects beyond the Hartree-Fock approximation (see, e.g., Ref.~[\onlinecite{GFG_GV}]), so we expect that this will not change qualitatively the critical $r_s$ for the ferromagnetic transition. Nevertheless, we cannot exclude the possibility that SDW states may appear in the phase diagram and this is a promising area for future research.

\subsection{Total energy}

We now consider the system without an electric field. We also set the disorder potential equal to zero which is permissible as long as localization effects are negligible. We expect this to be the case in a system with strong spin-orbit interactions, in which weak antilocalization rather than weak localization occurs at larger disorder concentrations. In the statically screened Hartree-Fock approximation, the exchange energy of the system can be written as
\begin{equation}\label{exchangestart}
E_\mathrm{ex} = - \frac{1}{A} \sum_{\bm{k} \ne \bm{k}'} 
\frac{\text{Tr} [e^{2} f_{\bm{k}} f_{\bm{k}'}]}
{2 \varepsilon_{r} \varepsilon_{0} (k_\mathrm{TF} + | \bm{k} - \bm{k}'|)}\, ,
\end{equation}
where $f_{\bm k}$ is the single-particle spin density matrix. The total energy of the system becomes $E_\mathrm{tot} = \trace [f_{\bm{k}} H_{0\bm{k}}] + E_\mathrm{ex}$. For convenience in the following numerical simulation the exchange energy (\ref{exchangestart}) may be rewritten as
\begin{equation}\label{exchangecode}
E_\mathrm{ex} = - \frac{1}{A} \sum_{\bm{k} \ne \bm{k}'}
\frac{e^{2} [\bm{s}_{\bm{k}}\cdot \bm{s}_{\bm{k}'} + n_{\bm{k}} \, n_{\bm{k}'}]}
{4 \, \varepsilon_{r} \varepsilon_{0} (k_\mathrm{TF} + | \bm{k} - \bm{k}'|)}\, ,
\end{equation}
where $n_{\bm{k}} = (1/2) \, \trace f_{\bm k}$ and $\bm{s}_{\bm{k}} = (1/2) \, \trace ({\bm \sigma} f_{\bm k})$ are the electron's occupation number
and net spin polarization at $\bm{k}$, respectively,
so the spin structure in Eq.~(\ref{exchangestart}) is replaced by the vector product
of spin polarizations.
Finally, the total energy becomes
\begin{equation}\label{Etot_exp}
E_\mathrm{tot} = \sum_{\bm{k}} \bigg[ \frac{\hbar^{2} k^{2}}{2m} n_{\bm{k}}
+ \alpha \, \bm{s}_{\bm{k}} \cdot ( \hat{\bm{k}} \times \hat{\bm{z}} ) \bigg] + E_\mathrm{ex}\, ,
\end{equation}
which will be a key variable in the following numerical simulations.

\subsection{Numerical procedure}

We wish to minimize $E_\mathrm{tot}$ with respect to the occupation numbers in momentum space and the spin directions.  The variables in Eq.~(\ref{Etot_exp}) are classical so we can use a classical Monte Carlo simulation to find the minimum-energy configuration. This is a significant advantage of the Hartree-Fock approximation. 
In the future spin-orbit-coupled systems should also provide a fruitful area for the Quantum Monte Carlo method. Some efforts have already been made in this direction, but have so far focused on the paramagnetic states \cite{PhysRevB.80.125306}.

For our classical Monte Carlo simulations, we discretize the reciprocal space
a simple equidistant mesh in $x$ and $y$ direction,
and replace the $\bm{k}$-integral by a summation over $N_\mathrm{dis}$ discrete wavevectors.
Each mesh point $\bm{k}$ in reciprocal space is characterized by two variables,
the occupation number $n_{\bm{k}}$ and the embodied spin direction $\bm{s}_{\bm{k}}$.  The
occupation numbers $n_{\bm{k}}$ can be 0, 1, and 2, which indicates an empty site, single occupancy, 
and double occupancy, respectively.
If $n_{\bm{k}} = 0$ or $2$, we have $\bm{s}_{\bm{k}} = 0$.  If $n_{\bm{k}} = 1$, the spin direction $\bm{s}_{\bm{k}}$ is a unit vector that is free to rotate in three dimensions (3D),
it is therefore characterized by two angles, the polar angle $\theta$ and the azimuthal angle $\phi$.
Note that this search method is limited to Slater determinants of momentum eigenstates,
so it does not find all candidate ground states.
For example, spin-density wave states are outside the search space.

The choice of the discretization number $N_\mathrm{dis}$ of the Fermi surfaces in the Monte Carlo simulations
was determined by running-time limitations and the desire to minimize numerical errors. The running times of the simulations increase quadratically with $N_\mathrm{dis}$,
so a reasonably small $N_\mathrm{dis} < 1000 $ is employed.
In order to compare the energies quantitatively, the value of $N_\mathrm{dis}$ was set to be fixed during all simulations,
but it was increased when necessary to identify phase boundaries.
In order to control the numerical error, there will be a lower limit for $N_\mathrm{dis}$.
Also, in determining the phases at small Rashba strength [$\tilde{\alpha} < 0.1$,
see Eq.~(\ref{alphatilde}) below for the definition of $\tilde{\alpha}$],
we always need to increase $N_\mathrm{dis}$ for more accurate Fermi surface structures.

In the following, we will assume that the screening effect is negligible due to the low electron density,
so $k_\mathrm{TF} = 0$.
The divergence of $V_{\bm q}$ does not cause any difficulty since the ${\bm q}=0$ term is absent from the discrete summations, due to the neutralizing background \cite{Vignale.05}.

The Monte Carlo simulation utilizes random numbers to decide
the evolution of the system status and calculate the averaged value of observables.
In our case, the acceptance criterion is $\exp[ -(E_{n+1} - E_{n} )/ k_\mathrm{B} T] > \omega$,
where $E$ is the total energy, $k_\mathrm{B}$ is the Boltzmann constant,
$n$ is the step number and $\omega$ is a random number from 0 to 1.
In the following, we will set $T = 0$,
which gives a ``greedy algorithm'' that only picks a lower energy state in every step,
although the $T > 0$ case will generally give information about critical temperatures of these ferromagnetic transitions.
In each step, we will allow two types of trial changes: the spin direction and the occupation number changes.
The spin direction change is quite straightforward, so we make it at first.
In the occupation number update we move one electron from a random occupied site to
another site that is not fully occupied $(n_{\bm{k}} < 2)$.
Note that the choice of the receiving site is random in the reciprocal space but,
to improve the efficiency of the algorithm, we assign a higher probability to the sites around the initial site.
All the other available sites in the reciprocal space can still be reached, although with lower probability. When the receiving site was previously empty $(n_{\bm{k}} = 0)$,
we need to transfer the spin of the previous site to the new one allowing random spin rotations. We parameterize the spin direction in terms of the Euler angles with $\theta$ the polar angle and $\phi$ the azimuthal angle. We set a maximum change for both $\theta$ and $\phi$, while the actual changes are evenly selected between zero and the corresponding maxima. We first change the azimuthal angle $\phi$ ([0, $2\pi$]) and then the polar angle $\theta$ ([0, $\pi$]). For the special case in which the receiving site is already singly occupied, $(n_{\bm{k}} = 1)$, the Pauli exclusion principle requires that the transferred electron has the opposite spin direction to the electron currently occupying the receiving site.

\section{Numerical Results}
\label{sec:NumRes}

We plot the phase diagram in Fig.~\ref{MonteCarlo_final}, as a function of the dimensionless variables $r_{s}$ and
\begin{equation}\label{alphatilde}
\tilde{\alpha} = \frac{m \alpha}{\hbar^{2} \sqrt{\pi n}}
= \frac{4\pi\varepsilon_0}{e^2}\,\varepsilon_r\, r_s\, \alpha
\equiv 0.07\,\varepsilon_r\, r_s\, \alpha [\mathrm{eV\,\AA}]\, ,
\end{equation}
where $r_{s}$ was defined in Eq.\ (\ref{eq:Wigner-Seitz}) while $\tilde{\alpha}$ is a measure of spin-orbit coupling relative to the kinetic energy.
We use $E_{kin}$ as the common base to characterize the strength of both Coulomb interaction and the spin-orbit coupling.

There are four different phases in Fig.~\ref{MonteCarlo_final}: FL1, FL2, OP, and IP phases,
which will be discussed in detail in the following subsections.

\begin{figure}[b]
\begin{center}
\includegraphics[width=0.8\columnwidth]{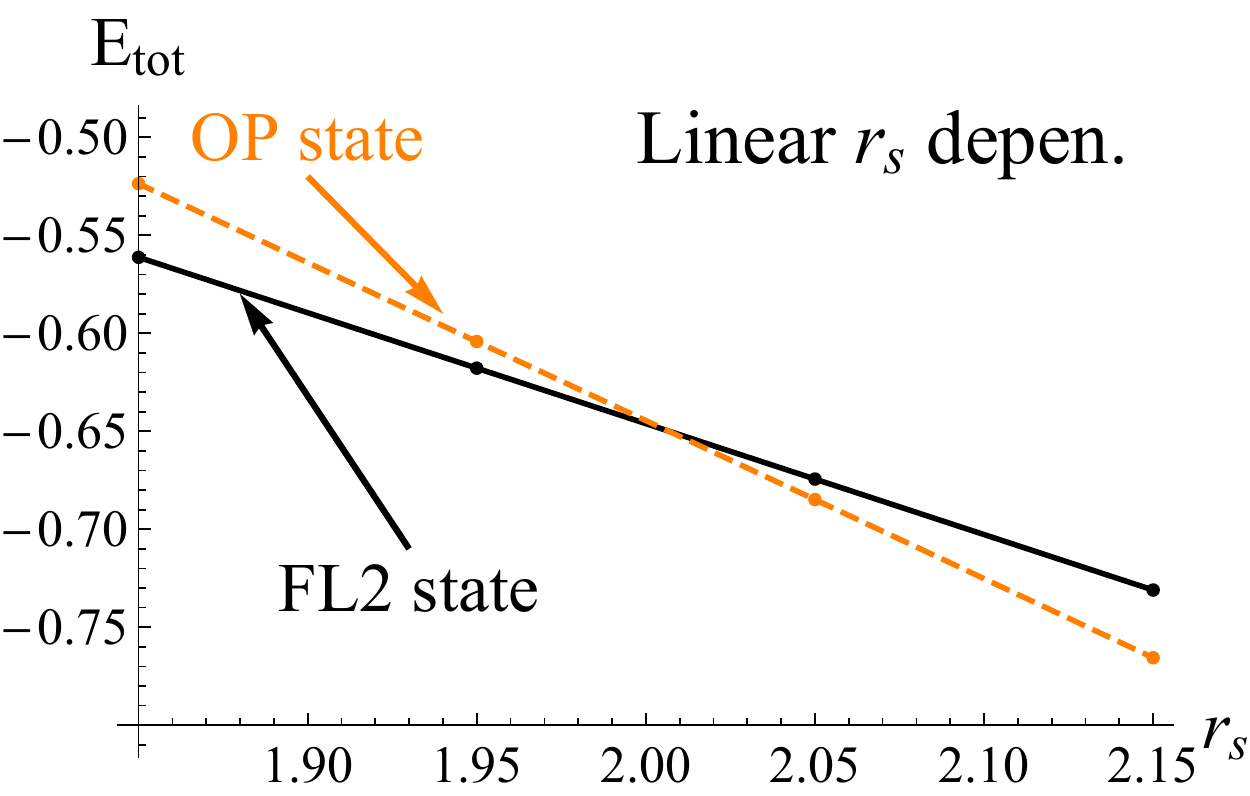}
\caption{The total energy $E_{\text{tot}}$ plots of the FL2 (black solid line) and OP (orange dashed line) states v.s.~$r_{s}$
at $\tilde{\alpha}= 0.12$.
The units of the energy are $N_{e} \hbar^{2} k_\mathrm{F}^{2} / 2m$,
where $N_{e} = n A$ is the total electron number.
We can see the linear $r_{s}$ dependence of the total energies when $r_{s}$ is close to phase boundaries,
which allows us to use the linear fitting to determine transition points.}
\label{Total_energy}
\end{center}
\end{figure}

\subsection{Fermi liquid phases}

The FL1 and FL2 phases are the conventional Fermi liquid (FL) states with one and
two occupied spin subband, respectively. The only effect of the exchange
interaction is to renormalize upwards the strength of the Rashba term
\cite{PhysRevB.85.075321, PhysRevA.90.043614, gfg_proc05, PhysRevB.75.155305}.
There is no net spin polarization. The phase boundary
separating FL1 and FL2 is well described by the (non-interacting) critical
density equation $n_{c} = m^{2}\alpha ^{2} / \pi \hbar ^{4}$
as noted above, which would give a horizontal boundary $\tilde{\alpha}_c=1.$
The exact boundary after considered the exchange interactions is:
\begin{equation}
\tilde{\alpha}_c=  1+\frac{\pi-1-2\mathcal{K}}{2\pi}r_s,
\end{equation}
where $\mathcal{K}\simeq 0.916$ is the Catalan's constant.
The small upward
slope is an indication that the interaction slightly favors the FL2 phase,
due to the effect of the $n_{\bm{k}}~n_{\bm{k}'}$ term near
$\bm{k}=\bm{k}' =0$ in Eq. (\ref{exchangecode}). The spin-orbit energy vanishes in
1st-order perturbation theory in the FL2 phase.

The quadratic dependence on spin-orbit-coupling strength for the total energy of FL phases is expected because,
when $\alpha$ changes sign, there will be no energy change at all.
The total energy of the FL2 state at $\tilde{\alpha} = 0.12$
is plotted in Fig.~\ref{Total_energy},
where the linear $r_{s}$ dependence is expected if $| r_{s} - r_{s}^\mathrm{tp}| \sim 0.1$
and $r_{s}^\mathrm{tp}$ is any transition points.

\subsection{OP phase}\label{OPsection}

\begin{figure}[b]
\begin{center}
\includegraphics[width=\columnwidth]{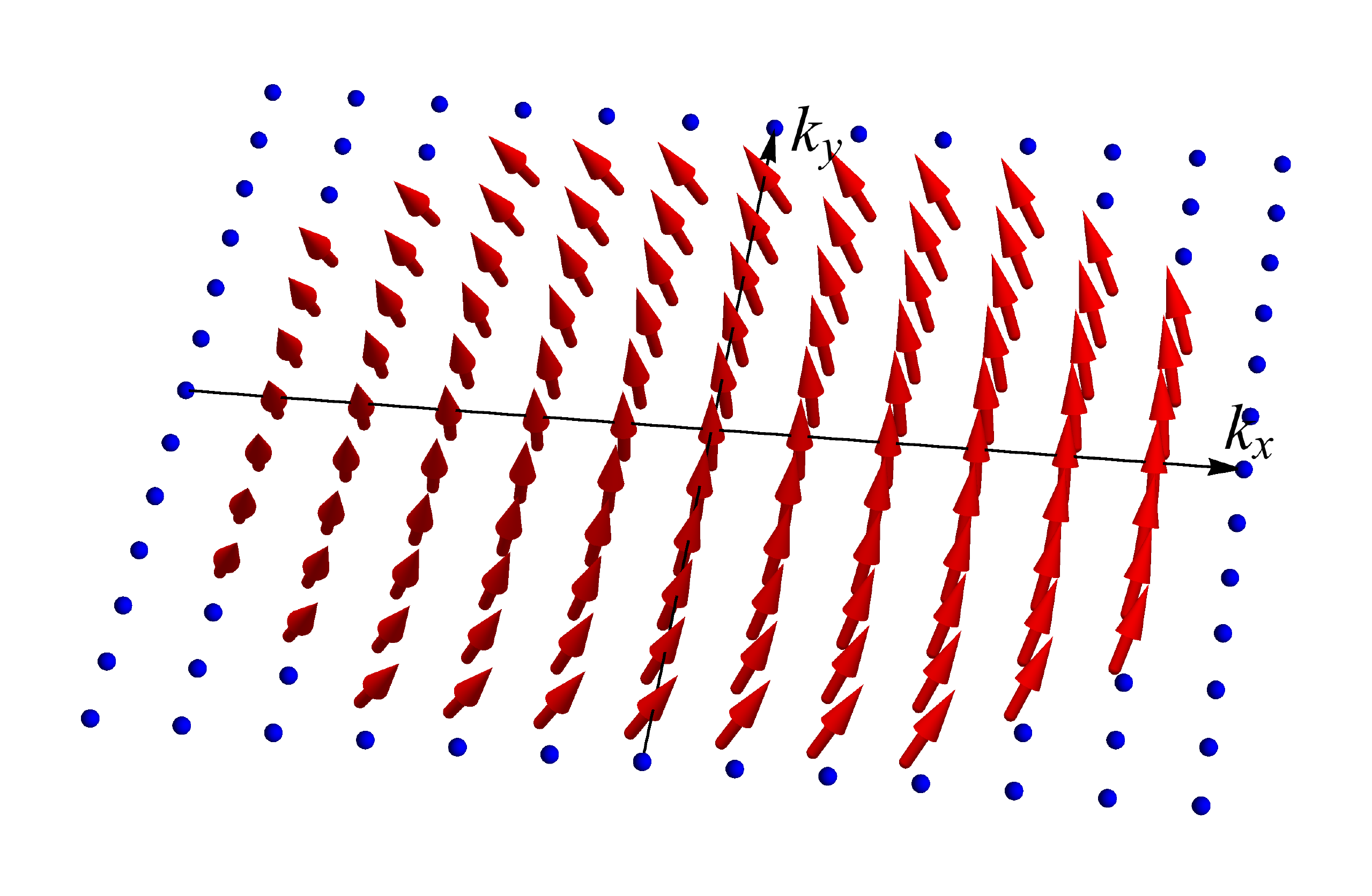}
\caption{The OP phase at $\tilde{\alpha}=0.3$ and $r_{s} = 2.02$ in a 3D view.
The OP phase comprises of a single band with circular Fermi surface and nontrivial out-of-plane spin polarization.
Its in-plane spin polarization cancels out after summing over all the occupied states.}
\label{OP_state}
\end{center}
\end{figure}

As $r_{s}$ increases, the interaction becomes more effective, producing a
tendency towards ferromagnetism. When $\alpha =0$, there is the classic
Bloch transition that occurs at 
\[
r_{s}^{*}=\frac{3\sqrt{2}\pi }{16\left( \sqrt{2}-1\right) }\approx 2.011 
\]
to a ferromagnetic state with magnetization along an arbitrary direction.
The numerical calculation is in excellent agreement with this analytical
result, see Fig.~\ref{MonteCarlo_final}.

When $\alpha $ is finite, then the ferromagnetic phase is modified to one that
we refer to as the OP phase. The spins have a $z$ component and a component along
the effective field due to the Rashba coupling. Thus at small $ k $
they point nearly along the $z$-direction,
but as $k$ increases they follow the spin orbit-induced
field. This is shown in Fig. \ref{OP_state}. This spin structure was first pointed
out in Refs.~\onlinecite{Juri2005, gfg_proc05}.
The underlying physical implication of the spin structure of the OP state is
the competition between the exchange interaction and the spin-orbit coupling.
The exchange interaction favors uniform alignment of all spins, while the spin-orbit coupling favors alignment of spins
following the local fields.

The transition from FL2 to OP is first-order, as we can see in Fig.~\ref{Total_energy},
so the boundary is given by the equation $E_\mathrm{FL2}\left( r_{s},\alpha \right) -E_\mathrm{OP}\left( r_{s},\alpha
\right) =0$, in an obvious notation. The effect of $\alpha $ on the OP
energy is quadratic. This then implies that the phase boundary
between the FL2 phase and the OP phase is vertical at $\tilde{\alpha }%
=0, $ since the boundary equation reduces to $r_{s}\left( \alpha \right)
=r_{s}\left( \alpha =0\right) +r_{s}^{\prime \prime }\alpha ^{2}/2$, and
then $\tilde{\alpha }\sim \left\vert r_{s}\left( \tilde{\alpha }%
\right) -r_{s}\left( \tilde{\alpha }=0\right) \right\vert ^{1/2}$. The
quadratic coefficient $r_{s}^{\prime \prime }$ is slightly negative,
favoring the OP phase.

\begin{figure}[b]
\begin{center}
\includegraphics[width=0.85\columnwidth]{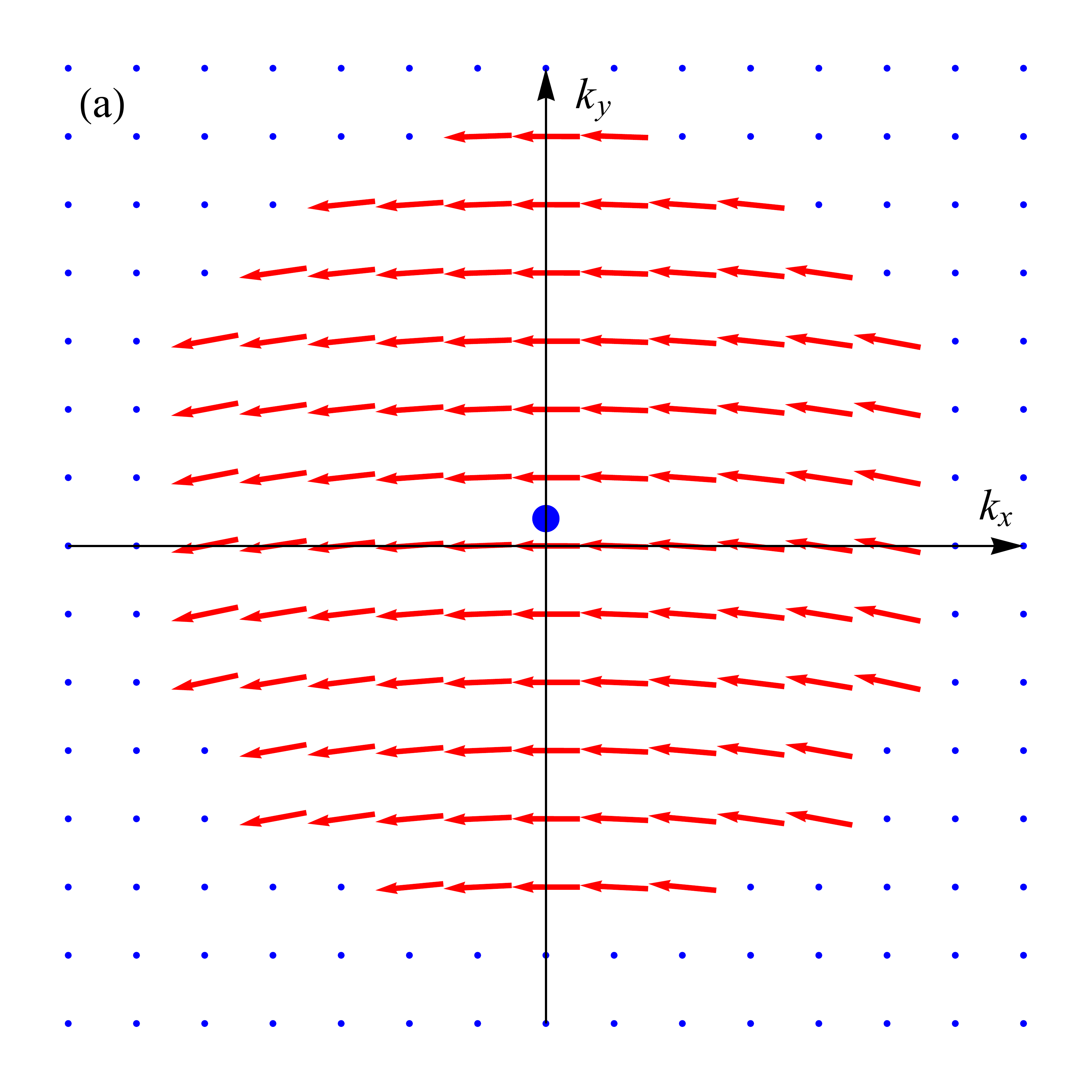}\\[2ex]
\includegraphics[width=1.\columnwidth]{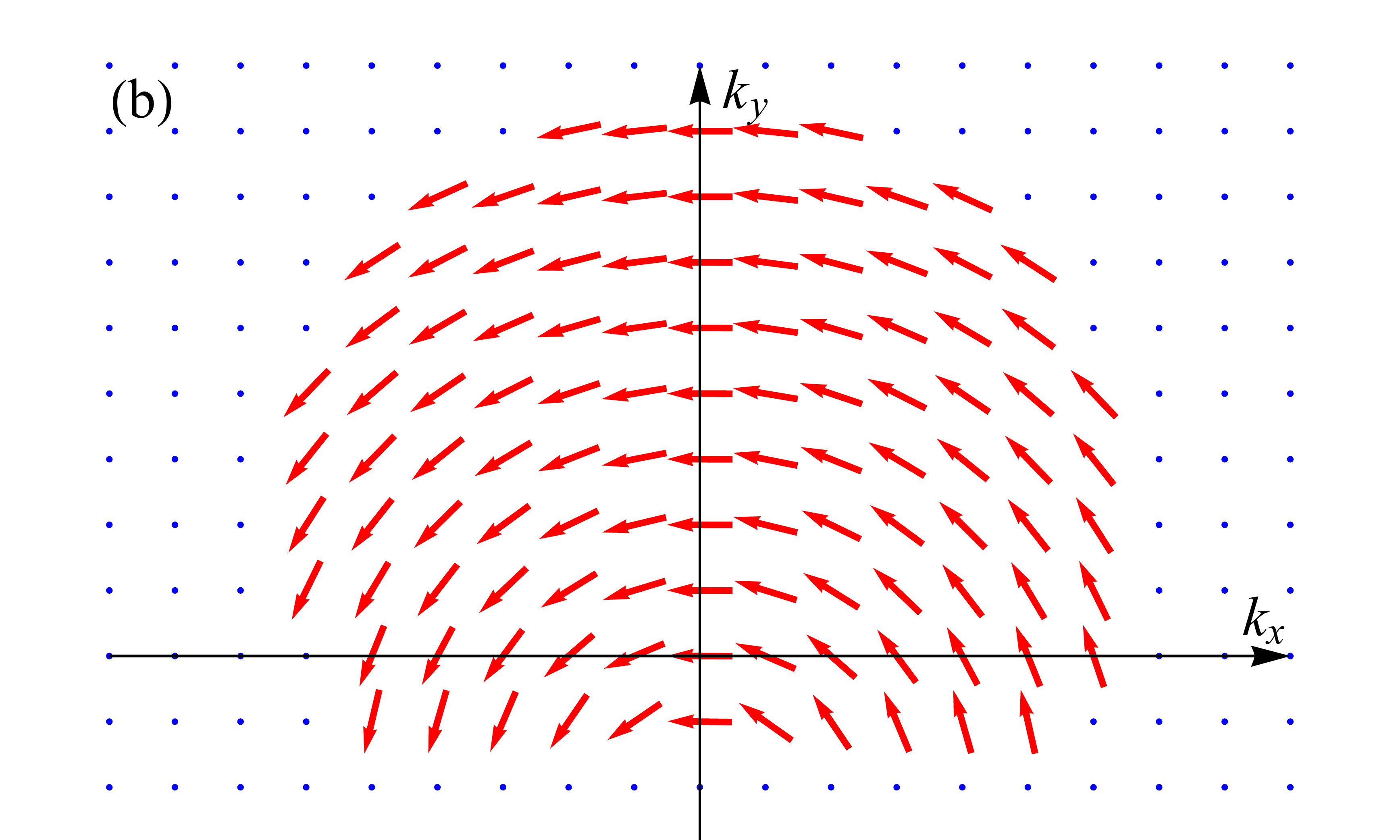}
\caption{\textbf{(a)} and \textbf{(b)} The IP phase at $\tilde{\alpha} = 0.12$ and $r_{s} = 2.25$,
and $\tilde{\alpha} = 1.13$ and $r_{s} = 2.12$.
In (a), the ratio $\tilde{\alpha} / r_{s} \sim 0.1$ is small and the Fermi surface is roughly symmetric with spins
almost parallel aligned.
In (b), the ratio $\tilde{\alpha} / r_{s} \sim 1$ is large and the Fermi surface is deformed into a ``heart'' shape.
The blue dot in (a) indicates the center of the displaced Fermi surface.
The spin texture in (b) follows the Fermi surface deformation, and the spins are winding around a center below the $x$ axis.}
\label{IP_plots}
\end{center}
\end{figure}

\subsection{IP phase}

The right half of the phase diagram in Fig.~\ref{MonteCarlo_final} is the IP phase,
which is the main finding of this paper.
The key feature of the IP phase is the spin polarization is completely in-plane, see Fig.~\ref{IP_plots}.
Compared with the OP state, the IP state does not have any symmetry on the Fermi surface,
though both of them only have a single band.
The spin texture of the IP state is also exotic and depends on the form of the Fermi surface.
When $\tilde{\alpha} / r_{s}$ is small $(\sim 0.1)$,
the Fermi surface is roughly a circle and all spin are almost parallel aligned, as shown in Fig.~\ref{IP_plots}(a).
In the limit of $r_{s} \to \infty$, the Fermi surface becomes a rigid circle,
and the displacement of the Fermi surface is exactly $\Delta k = k_{c}$,
as required by the zero current condition.
Here the displacement can be along any in-plane directions and the spin polarization is always perpendicular to
the displacement vector.
In the large $\tilde{\alpha} / r_{s} (\sim 1)$ case, the Fermi surface becomes ``heart'' like, see Fig.~\ref{IP_plots}(b),
and the spin texture takes on a complex form. The shape of Fig.~\ref{IP_plots}(b) is reminiscent of the Pomeranchuk instability. 

At $\tilde{\alpha }=0$, the direction of the magnetization is arbitrary
for all $r_{s}$. However, any small field destroys this isotropy, and the
spin-orbit field can play this role. This is what happens at the point
where the OP-IP phase boundary intersects the $\tilde{\alpha }=0$ axis
in Fig.~\ref{MonteCarlo_final}. At any finite $\tilde{\alpha }$ the symmetry is broken and
we have either the OP or the IP phase, depending on the value of $r_{s}$.

\subsection{Other Phases}

We recall that, in the non-interacting Rashba electron system, for $n>n_{c}$
there are two circular Fermi surfaces, while for $n<n_{c}$ there is a single
disc-shaped Fermi surface. It was recently shown using the Hartree-Fock
approximation that, for $n<n_{c}$, this disk can break up into two pockets
with either ferromagnetic or Neel order \cite{Ruhman_PRB14}. We note that in
a semiconductor $n_{c}$ is extremely small and is nearly impossible to
realize experimentally, though it may be achievable in cold atom setups. \
Hence we have not attempted to locate these phases.

\section{Magnitude of the Fermi surface shift}

Because of the complexity of the ${\bm k}$-space occupation in the IP phase,
numerical calculations are required to fully understand it. However, it is
important from several points of view to have a qualitative understanding of
the rough size of the magnitude of the shift of the Fermi surface in ${\bm k}
$-space. So here we present asymptotic analyses to give semi-quantitative
estimates of the shift in different parts of the phase diagram. We define
the average shift ${\bm q}$ by the equation%
\[
{\bm q}=\frac{\int d^{2}k~{\bm k}~n_{\bm k}}{\int d^{2}k~n_{\bm k}}
\]%
where $n_{\bm k}$ is the occupation of state $\vec{k}$ summed over spin. \
The direction of ${\bm q}$ is not fixed by the Hamiltonian since the IP
phase is the result of a spontaneously broken rotational symmetry. \ The
magnitude is of great interest, since the larger $q=\left\vert {\bm q}%
\right\vert $ is, the easier it will be to detect experimentally.

For a rough estimate of $q,$ we only need to understand the $q$-dependences
of the various contributions to the total energy. \ The kinetic energy is
the simplest. \ For parabolic bands we have for each ${\bm k}$-state that a
shift by ${\bm q}$ increases the energy from $\hbar ^{2}k^{2}/2m$ to $\hbar
^{2}\left( {\bm k}+{\bm q}\right) ^{2}/2m.$ \ On integration over $\vec{k}$
the cross term approximately cancels and we find that the dependence of the
kinetic energy on $q$ has the form $\hbar ^{2}q^{2}n/2m.$ \ 

At small $q$ ($q<<k_{F}$) the spin-orbit energy is quadratic in $q$ and we
write it as $-a_{so}nq^{2}$ but at large $q$ ($q>>k_{F}$) it is linear since
the spins follow the effective field in that case and we have $-\alpha nq.$
At small $q$ ($q<<k_{F}$) the exchange energy is also quadratic in $q:$ $%
-a_{ex}nq^{2}$ but at large $q$ ($q>>k_{F}$) it saturates since the spin
polarization is complete and the asymptotic exchange energy density is $%
E_{ex}/A$. \  

\subsection{Near the FL1-IP boundary}

This transition is continuous and $q$ may be regarded as the order parameter
of the transition: its appearance marks the onset of the spontaneous
breaking of the rotational symmetry and a Ginzburg-Landau analysis is
appropriate. \ For small $q$ the difference in energy of the two phases is 
\[
\frac{1}{A}(E_{FL1}-E_{IP})=\left( \hbar ^{2}n/2m-a_{so}-a_{ex}\right)
q^{2}+O\left( q^{4}\right) ,\text{ for }q<<k_{F}.
\]%
The transition is signaled as usual by the change in sign of the quantity in
parentheses. \ $a_{so}$ increases with $\alpha $ and $a_{xe}$ with $r_{s},$
giving the rough shape of the phase boundary. \ $q$ grows continuously from
zero. \ Interestingly, there is no identifiable large $q^{4}$ term in this
analysis, suggesting that $q$ grows very rapidly as we move away from the
phase boundary. \ 

\subsection{Near the OP-IP phase boundary}

This is a first-order transition, so the energies of the two phases need to
be estimated separately. \ We have 
\begin{equation}
\frac{E_{\mathrm{IP}}\left( q\right) }{A}=\frac{\hbar ^{2}\pi n^{2}}{m}+%
\frac{\hbar ^{2}nq^{2}}{2m}-a_{so}nq^{2}-a_{ex}nq^{2},  \label{IPOP_ZEROTH}
\end{equation}%
For the OP phase 
\[
\frac{E_{\mathrm{OP}}}{A}=\frac{\hbar ^{2}\pi n^{2}}{m}-\frac{2\pi
\varepsilon _{r}\varepsilon _{0}\alpha ^{2}\left( \pi n\right) ^{3/2}}{%
3e^{2}\left( 1-C\right) },
\]%
where $C$ is a pure number that describes the spin-orbit energy in the OP
spin texture.

Setting $E_{\mathrm{OP}}=E_{\mathrm{IP}}$ we see that $q$ is proportional to 
$\alpha $ along the IP side of the phase boundary. \ Thus $q$ vanishes on
the horizontal axis in Fig. 1 and grows linearly along it. \ However, at the
upper end of the boundary where we come to the FL1-IP boundary $q$ must
again vanish. \ Hence we expect that $q$ will be small along the OP-IP
boundary and this expectation is borne out by Fig. 4(a). \  

\subsection{Deep in the IP phase}

At large $q$ we have%
\[
\frac{1}{A}E_{IP}=\frac{\hbar ^{2}n}{2m}~q^{2}-\alpha nq-\frac{1}{A}E_{ex},%
\text{ for }q>>k_{F},
\]%
with an equilibrium 
\[
q=\frac{m\alpha }{\hbar ^{2}}.
\]%
\ Thus the magnitude of the shift is determined by the competition between
spin-orbit energy and kinetic energy since the exchange energy is saturated.
\ We have returned to the non-interacting case, since this value of $q$ is
just same as the shift in the minimum of the non-interacting dispersion
relation caused by Rashba spin-orbit coupling. \ This shift can be large, as
seen in Fig. 4(b). \ The equation also predicts that the shift is
approximately independent of $r_{s}$ deep in the IP phase. \ We have
verified this in the numerical simulations, though we do not present a
detailed analysis here.

\section{The limit of small spin-orbit coupling} \label{sec:small_alpha}

Although for general parameters the ground state has complex features which can only be characterized numerically, an analytical treatment can be developed in the regime of small spin-orbit coupling. This treatment, which is complementary to the solution of the HF problem by the classical Monte Carlo minimization, is described in this section. 

Since only one spin band is occupied in the ferromagnets, we consider a state described by $f_{\bm k}=n_{\bm{k}}(1+  \bm{\theta}_{\bm{k}}\cdot \bm{\sigma} )/2$ where $n_{\bm{k}}=0,1$ and $\bm{\theta}_{\bm{k}}$ is a unit vector. $n_{\bm{k}}$ and $\bm{\theta}_{\bm{k}}$ have only small corrections (of order $\alpha$) from their unperturbed values:
\begin{equation}
n_{\bm{k}}=\Theta\left(\sqrt{2}k_F	-k \right)\equiv n_0(k), \qquad \bm{\theta}_{\bm{k}}=\bm{\theta}_0,
\end{equation}
where $\bm{\theta}_0$ gives the polarization direction of the ferromagnet. Our analysis is based on the single-particle mean-field Hamiltonian
\begin{equation}
H_{0\bm{k}}-\mathcal{B}_{\bm{k}} = \frac{\hbar^{2} k^{2}}{2 m} + \alpha\,\bm{\sigma}\cdot
(\bm{k} \times \hat{\bm{z}})- \int \frac{\mathrm{d} \bm{k} }{(2\pi)^{2}} V_{\bm{k} - \bm{k}'} f_{\bm{k}'},
\end{equation}
which should be solved self-consistently for the HF ground state. This implies the following condition
\begin{equation}\label{theta_eq}
\bm{\theta}_{\bm{k}}=\frac{1}{C_{\bm{k}} }\left[\alpha (\hat{\bm{z}}\times \bm{k} )  + \int \frac{\mathrm{d} \bm{k}' }{(2\pi)^{2}} V_{\bm{k} - \bm{k}'} \frac{n_{\bm{k}'}}{2} \bm{\theta}_{\bm{k}'}\right].
\end{equation}
The physical meaning of Eq.~(\ref{theta_eq}) is that the spin direction $\bm{\theta}_{\bm{k}}$ must be parallel to the total effective field at $\bm{k}$, which is the sum of the spin-orbit and exchange fields (the two terms in the square parenthesis). $C_{\bm{k}}$ is a scalar insuring that $\bm{\theta}_{\bm{k}}$ is a unit vector. In the unperturbed case $C_{\bm{k}}$ is isotropic:
\begin{equation}
C_{\bm{k}} = \int \frac{\mathrm{d} \bm{k}' }{(2\pi)^{2}} V_{\bm{k} - \bm{k}'} \frac{n_0(k')}{2} \equiv C_0(k).
\end{equation}

Once suitable spin directions are found, the single-particle energies are given by:
\begin{align}\label{ek}
\epsilon_{\bm{k}} = & \frac{\hbar^{2} k^{2}}{2 m} + \alpha\,
(\bm{k} \times \hat{\bm{z}})\cdot \bm{\theta}_{\bm{k}} \nonumber \\
& - \int \frac{\mathrm{d} \bm{k}' }{(2\pi)^{2}} V_{\bm{k} - \bm{k}'} n_{\bm{k}'}\frac{1+  \bm{\theta}_{\bm{k}'}\cdot \bm{\theta}_{\bm{k}}}{2}.
\end{align}
For the ground state, $\epsilon_{\bm{k}}$ must be constant on the Fermi surface. At $\alpha=0$, this is automatically true because $\epsilon_{\bm{k}}$ is a function of $k$. However, both $n_{\bm{k}}$ and $\bm{\theta}_{\bm{k}}$ are generally anisotropic at finite $\alpha$, which makes the requirement of a constant Fermi energy non-trivial.

\subsection{Spin texture}

\begin{figure}
\begin{center}
\includegraphics[width=0.8\columnwidth]{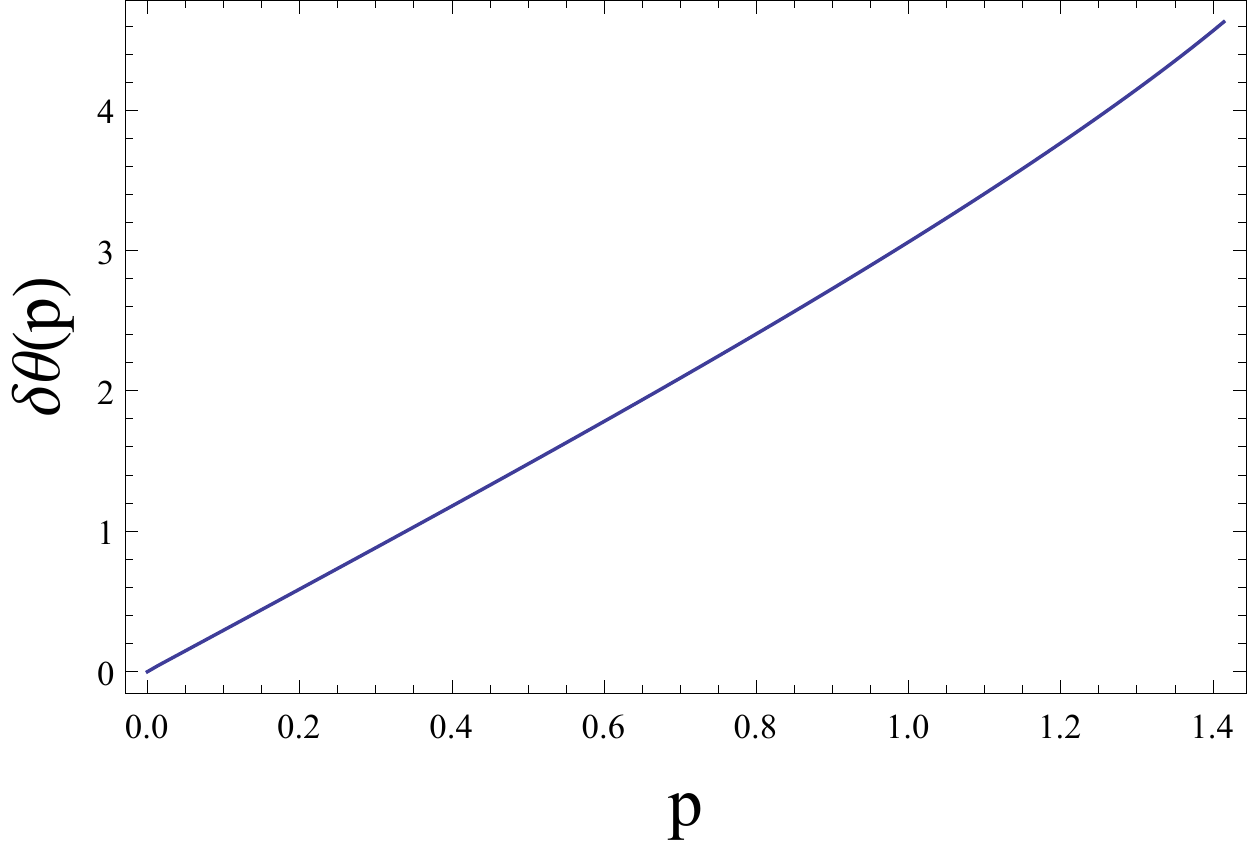}
\caption{Solution of Eq.~(\ref{integral_eq}). The function $\delta\theta(p)$ determines the spin texture of the ferromagnetic states at small $\alpha$, see Eq.~(\ref{thetak}), as well as the critical point $r_s^{**}$, see Eq.~(\ref{rstar}).}
\label{dtheta}
\end{center}
\end{figure}

We start from the corrections to $\bm{\theta}_{\bm{k}}$ which, to lowest order in $\alpha$, are orthogonal to $\bm{\theta}_0$. In the following, we assume without loss of generality $\bm{\theta}_0\cdot \hat{\bm{y}} =0$. Discarding $\alpha^2$ corrections, Eq.~(\ref{theta_eq}) gives:
\begin{equation}\label{dtheta1}
\bm{\theta}_{\bm{k}}\cdot \hat{\bm y} =\frac{1}{C_0(k)}\left[\alpha k_x + \int \frac{\mathrm{d} \bm{k}' }{(2\pi)^{2}} V_{\bm{k} - \bm{k}'} \frac{n_0(k')}{2} \bm{\theta}_{\bm{k}'}\cdot \hat{\bm y}\right].
\end{equation}
Here, the angular dependence of $\bm{\theta}_{\bm{k}}\cdot \hat{\bm y}$ is determined by the perturbation $\alpha k_x$. Therefore, we define $\bm{\theta}_{\bm{k}}\cdot \hat{\bm y} = \tilde{\alpha} r_s^{-1} \delta\theta(k/k_F) k_x/k$
and transform Eq.~(\ref{dtheta1}) in a one-dimensional integral equation for $\delta\theta(p)$:
\begin{equation}\label{integral_eq}
\delta\theta(p) =\frac{4\pi p + \int_D d\bm{p}' \left(\hat{\bm{p}}\cdot \hat {\bm{p}}' |\bm{p}-\bm{p}'|^{-1}\right) \delta\theta(p') }{ \int_D d\bm{p}'  |\bm{p}-\bm{p}'|^{-1}},
\end{equation} 
where $\bm{p}= \bm{k}/k_F$ is a dimensionless vector, with direction $\hat{\bm{p}}=\bm{p}/p$. The integration domain $D$ corresponds to $n_0(k)$ and is a disk with radius $\sqrt{2}$. The solution of Eq.~(\ref{integral_eq}) is found numerically and is shown in Fig.~\ref{dtheta}.

The analysis of the component along $(\bm{\theta}_0 \times \hat{\bm{y}})$ is similar to $\bm{\theta}_{\bm{k}}\cdot \hat{\bm y} $. It turns out that the solution of Eq.~(\ref{integral_eq}) fully characterizes the spin texture of the perturbed ferromagnet which has the following form:
\begin{equation}\label{thetak}
\bm{\theta}_{\bm{k}} \simeq \bm{\theta}_0 + \frac{\tilde{\alpha}}{r_s} \delta\theta(k/k_F) \frac{k_x \hat{\bm y}+ k_y(\bm{\theta}_0 \cdot \hat{\bm{z}}) (\bm{\theta}_0 \times \hat{\bm{y}})}{k}.
\end{equation}
For an OP state, with $\bm{\theta}_0 =  \hat{\bm{z}}$, the second term of Eq.~(\ref{thetak}) becomes proportional to $\bm{k} \times \hat{\bm{z}}$, i.e., the perturbation has the same angular dependence of the spin-orbit interaction. On the other hand, for an IP state with $\bm{\theta}_0 =\hat{\bm{x}}$, $\bm{\theta}_{\bm{k}}$ becomes slightly canted in-plane, along the $\hat{\bm y}$ direction perpendicular to the initial polarization. 

\subsection{Displaced Fermi surface}

To complete our analysis of the ground state, we should find the effect of spin-orbit coupling on the occupation numbers $n_{\bm{k}}$. To this end we simplify Eq.~(\ref{ek}) by making use of Eq.~(\ref{thetak}):
\begin{equation}\label{ek2}
\epsilon_{\bm{k}} 
\simeq   \frac{\hbar^{2} k^{2}}{2 m} + \alpha 
(\bm{\theta}_{0} \cdot \hat{\bm{x}})k_y - \int \frac{\mathrm{d} \bm{k}' }{(2\pi)^{2}} V_{\bm{k} - \bm{k}'} n_{\bm{k}'}.
\end{equation}
Since in Eq.~(\ref{ek2}) the perturbation $ \alpha (\bm{\theta}_{0} \cdot \hat{\bm{x}})k_y $ is not constant on the unperturbed Fermi surface $k=\sqrt{2}k_F$, it drives a change of $n_{\bm{k}}$. It is easily checked that modifying the Fermi surface as follows:
\begin{equation}\label{kshift}
\bm{k} \to \bm{k}-\frac{m\alpha}{\hbar^2}  (\bm{\theta}_{0} \cdot \hat{\bm{x}}) \hat{\bm{y}}, \quad  {\rm (with}~k= \sqrt{2} k_F{\rm)},
\end{equation}
yields a constant single-particle energy, independent of the direction $\hat{\bm{k}}$. Thus, Eq.~(\ref{kshift}) gives the desired change of the Fermi surface to first order in $\alpha$.

Equation~(\ref{kshift}) is a simple translation of the Fermi surface which does not affect the exchange contribution to $\epsilon_{\bm{k}} $ [i.e., the last integral of Eq.~(\ref{ek2}) is unchanged for a simultaneous shift of $\bm{k}$ and $n_{\bm{k}'}$]. Therefore, Eq.~(\ref{kshift}) is decided by the non-interacting part and can be interpreted on the basis of the single-particle velocity $v_y=\hbar k_y/m + \alpha \sigma_x /\hbar$. If we require that $\langle v_y \rangle =0$, we obtain $\langle k_y \rangle \simeq -\frac{m\alpha}{\hbar^2} (\bm{\theta}_{0} \cdot \hat{\bm{x}}) $ in agreement with Eq.~(\ref{kshift}). As expected, the Fermi surface is unchanged for a OP state ($\bm{\theta}_{0} = \hat{\bm{z}}$) and the maximum shift is obtained for the IP state ($\bm{\theta}_{0} = \hat{\bm{x}}$).

\subsection{Energy and phase boundaries}

Finally, we compute the total energy to lowest order in $\alpha$, which allows us to discuss the boundaries between different phases. To make use of the previous characterization of the ferromagnets, we can apply standard results of linear-response theory to the total Hamiltonian $H \equiv H_\alpha$. With a obvious notation (i.e., $R=H_\mathrm{SO}/\alpha$), we write:
\begin{equation}\label{Halpha}
H_\alpha =H_{0} + \alpha R,
\end{equation}
where the ground state $|\psi_\alpha \rangle$ gives the total energy $E_\alpha= \langle \psi_\alpha | H_{\alpha} |\psi_\alpha  \rangle $. The susceptibility $\chi_{RR}$ is defined by:
\begin{equation}
\langle \psi_\alpha | R |\psi_\alpha \rangle \simeq \chi_{RR} \alpha,
\end{equation}
and is related to $E_\alpha$ as follows \cite{Vignale.05}:
\begin{equation}
E_\alpha = E_0 + \frac12\chi_{RR} \alpha^2 =  E_0 + \frac12 \langle \psi_\alpha | \alpha R |\psi_\alpha  \rangle .
\end{equation}
The last equation is very convenient, because it expresses the change in total energy as one-half of the spin-orbit interaction energy. Since the spin-orbit interaction is already linear in $\alpha$, the first-order corrections to $|\psi_\alpha \rangle$ are sufficient to obtain the desired $\propto \alpha^2$ energy change. Explicitly:
\begin{align}\label{EN_alpha2}
& \frac{E_\alpha - E_0}{N_e}  = \frac{\alpha }{2n} \int \frac{d\bm{k}}{(2\pi)^2} n_{\bm{k}} \, \bm{\theta}_{\bm{k}} \cdot  (\bm{k}\times  \hat{\bm{z}}) \nonumber \\
& = -\frac{\alpha^2 m}{2\hbar^2} \left[ (\bm{\theta}_{0} \cdot \hat{\bm{x}})^2 
 + \frac{1+(\bm{\theta}_{0} \cdot \hat{\bm{z}})^2}{\sqrt{2}r_s} \int_0^{\sqrt{2}}\delta\theta(p)p^2 dp\right],
\end{align}
where in the second line we have used Eqs.~(\ref{thetak}) and (\ref{kshift}). Since $(\bm{\theta}_{0} \cdot \hat{\bm{x}})^2=1-(\bm{\theta}_{0} \cdot \hat{\bm{z}})^2 $, the above expression shows that the  minimum energy is attained by the OP phase at sufficiently large density $r_s < r_s^{**}$ (when the second term in the square parenthesis dominates) while for $r_s > r_s^{**}$ the ground state is in the IP phase. The critical density is obtained from $\delta\theta(p)$ by numerical integration:
\begin{equation}\label{rstar}
r_s^{**}=\frac{1}{\sqrt{2}} \int_0^{\sqrt{2}}\delta\theta(p)p^2 dp \simeq 2.21.
\end{equation}
Since $r_s^{**}$ is larger than $r_s^*\simeq 2.011$ (the value of the classical Bloch transition), there is an OP region between the high-density FL2 paramagnet and the low-density IP ferromagnet. Equation~(\ref{rstar}) is in good agreement with the direct numerical simulation based on the Monte Carlo method, see Fig.~\ref{MonteCarlo_final}.

The evaluation of the energy of the ferromagnetic states also allows us make more rigorous the discussion at the end of Sec.~\ref{OPsection}, about the boundary between the FL2 and OP phases. The OP state is a special case of Eq.~(\ref{EN_alpha2}):
\begin{equation}\label{EOP}
\frac{E_\mathrm{OP}}{N_e}= \frac{\hbar^2 k_F^2}{2 m}\left(2-\frac{16r_s}{3\pi}-\frac{r_s^{**}}{r_s}{\tilde\alpha}^2 \right),
\end{equation}
where we have substituted to $E_0$ the well-known energy of the $\alpha=0$ ferromagnet. For the paramagnetic phase we have \cite{PhysRevB.83.235309}:
\begin{equation}\label{EFL2}
\frac{E_\mathrm{FL2}}{N_e}= \frac{\hbar^2 k_F^2}{2 m}\left(1-\frac{8 \sqrt{2}r_s}{3\pi}-{\tilde\alpha}^2 \right),
\end{equation}
where the $\alpha^2$ correction is given by the non-interacting Hamiltonian. In fact, the exchange energy of the paramagnetic state (the second term in the parenthesis) is only modified by a term of order $\alpha^4 \ln\alpha$ \cite{PhysRevB.83.235308, PhysRevB.83.235309} which is negligible for the present discussion. Equating Eqs.~(\ref{EOP}) and (\ref{EFL2}) gives the small-$\alpha$ phase boundary:
\begin{equation}
\tilde{\alpha}= \sqrt{\frac{r_s(r_s^* - r_s)}{r_s^*(r_s^{**} - r_s)}} \simeq 2.24 \sqrt{r_s^* - r_s} ,
\end{equation}
where in the last step we used $r_s\simeq r_s^*$. We see that at finite $\alpha$ the phase bondary occurs at $r_s < r_s^*$, i.e., the presence of spin-orbit coupling slightly favors the formation of the OP phase. This conclusion is in agreement with the numerical phase diagram of Fig.~\ref{MonteCarlo_final}.

\section{Discussion}
\label{sec:disc}

The competition among the kinetic, interaction, and spin-orbit contributions to the electronic energy produces a rich variety of phases in the parameter space that varies the relative strengths of these energies.  We have identified 3 distinct phases: Fermi liquid (both FL1 and FL2), OP, and IP.  The transitions between these phases appear to be first-order in all cases. The IP phase in particular comprises a rich variety of spin textures that interpolate between the vortex-like structure induced by the spin-orbit field in momentum space and the ferromagnetic structure in the limit where kinetic energy is small.

The phase diagram, Fig.~\ref{MonteCarlo_final}, of the Rashba spin-orbit coupled system
contains a significant amount of information. The Fermi liquid phases, FL1 and FL2 phases, have been studied in the past. In particular, the FL1 phase is realized at $n\lesssim n_{c}$,
with occupation in the form of a ring.
For very low density ($n \ll n_{c}$) case, the FL1 phase will be a ring,
and then the strong exchange interaction would deform the FL1 phase into a
one-node ferromagnetic or two-nodes ``nematic'' state,
as shown in Ref.~\cite{Ruhman_PRB14}.
In contrast to previous studies, the IP ferromagnetic phase appears much more prominently in our phase diagram,
showing that the demanding condition $n<n_{c}$ is not necessary.
According to Eq.~(\ref{alphatilde}), the values of $\alpha$ required to enter the regime with a nontrivial interplay with the OP
phase are routinely achievable.

We expect that the phase diagram Fig.~\ref{MonteCarlo_final}
also applies to the linear Dresselhaus spin-orbit coupled systems \cite{PhysRev.100.580, Winkler03},
since the latter has the same energy spectrum as the one in the linear Rashba spin-orbit system.

\subsection{In-plane spin polarized phase}

Compared with Refs.~\cite{gfg_proc05, Juri_PRB08}, we introduce the asymmetric change of the Fermi surface
and discover the missing part of the phase diagram, the IP phase.
If the Fermi surface is restricted to be circular,
then the results of Refs.~\cite{gfg_proc05, Juri_PRB08} are easily explained as follows.
When the exchange interaction is small ($r_{s} < 2$),
the spin-orbit coupling plays the key role in the band structure
and is renormalised by the exchange interaction in a perturbative way.
When the exchange interaction is very strong ($r_{s} > 2$),
the spin alignment due to the exchange interaction
can only result in the out-of-plane spin polarization.
The spin directions are tilted in-plane, to form a spin-winding in momentum space which follows the non-interacting Bloch states. However, the spin-orbit interaction is still greatly penalized by the nearly parallel alignment.
If we allow asymmetric deformation of the Fermi surface, then the situation becomes complicated,
because the in-plane spin polarization state can be formed together with a Fermi surface displacements,
same as the electrical field case discussed in Sec.~\ref{sec:transport}.
In general, when $r_{s} \to \infty$, the state prefers the in-plane phase rather than the out-of-plane phase,
since the former one can lower the total energy by gaining a significant amount of spin-orbit energy with respect to the latter.
The introduction of the asymmetric deformation of the Fermi surface in the accessible $n> n_{c}$ regime
and the resulting in-plane spin states are the central finding of this paper.

\subsection{Absence of electrical current in the in-plane spin-polarized phase}

We would like to emphasize that, even though the Fermi surface is displaced from equilibrium, a simple standard argument demonstrates that there is no net charge current in the equilibrium system, as one expects from basic physical considerations. We note that this is also consistent with our finding that the charge conductivity is unaffected by the diverge in the spin polarization. Briefly, in the basis of eigenstates of the interacting system the expectation value of the current operator is simply the integral of the group velocity over reciprocal space. If the eigenenergies of the interacting system are denoted by $\varepsilon_{n{\bm k}}$, the net current density is:
\begin{equation}
\label{eq:charge-cur}
{\bm j} = - \frac{e}{\hbar} \int \frac{\mathrm{d}^2 k}{(2\pi)^2} \, n_{\bm k}\, \frac{\partial \varepsilon_{n \bm{k}}}{\partial \bm{k}},
\end{equation}
where the integral runs over all ${\bm k}$ and, as above, $n_{\bm k}$ is the occupation of each eigenstate. We take for concreteness the $x$-component of this equation. At $T=0$ one way to evaluate this is to use $n_{\bm k}$ to fix the limits of integration
\begin{equation}\label{jx}
\begin{array}{rl}
\displaystyle j_x = & \displaystyle - \frac{e}{\hbar} \int_{k_{Fy-}}^{k_{Fy+}} \frac{\mathrm{d}k_y}{2\pi} \int_{k_{Fx-}}^{k_{Fx+}} \frac{\mathrm{d}k_x}{2\pi} \, \frac{\partial \varepsilon_{n\bm{k}}}{\partial k_{x}} \\ [1ex]
= & \displaystyle - \frac{e}{h} \int_{k_{Fy-}}^{k_{Fy+}} \frac{\mathrm{d}k_y}{2\pi} \bigg[ \varepsilon_{n{\bm k}} \bigg]_{k_{Fx-}}^{k_{Fx+}}. 
\end{array}
\end{equation}
Here $k_{Fx-}$ and $k_{Fx+}$ represent the $x$-components of the Fermi wave vectors, with identical notation for the $y$-components. It is seen that $j_x$ vanishes identically since both $k_{Fx-}$ and $k_{Fx+}$ are on the Fermi surface, making the energies equal.

Alternatively, one can integrate Eq.~(\ref{eq:charge-cur}) by parts. Again, considering the $x$-component of this equation 
\begin{equation}
\begin{array}{rl}
\displaystyle j_x = & \displaystyle \frac{e}{\hbar} \int \frac{\mathrm{d}^2 k}{(2\pi)^2} \, \varepsilon_{n \bm{k}} \, \frac{\partial n_{\bm k}}{\partial k_x}.
\end{array}
\end{equation}
Here we note that $n_{\bm k}$ has two discontinuities as a function of $k_x$, one at $k_{Fx-}$ and one at $k_{Fx+}$. With this in mind we can write
\begin{equation}
\begin{array}{rl}
\displaystyle j_x = & \displaystyle \frac{e}{h} \int \frac{\mathrm{d} k_y}{2\pi} \, \varepsilon_{n \bm{k}} \, [\delta(k_x - k_{Fx-}) - \delta(k_x - k_{Fx+})] = 0,
\end{array}
\end{equation}
in agreement with the above.

\subsection{Symmetry considerations}

The in-plane spin polarized phase is accompanied by a sizable shift in the Fermi surface and consequently involves the creation of a spontaneous net spin-orbit effective field. This recalls electrically-induced spin polarization, which we recall occurs only in materials that are gyrotropic, meaning that light with left- and right-rotating elliptical polarizations can propagate at different speeds. Hence we expect in-plane spin-polarized phases with shifted Fermi surfaces to emerge in systems displaying electrically-induced spin polarizations. This argument proves that the in-plane spin-polarized phase is qualitatively different from ordinary Stoner ferromagnetism, which is not subject to these symmetry restrictions. Rather, this phase is reminiscent of the Pomeranchuk instability. 

Although a number of 3D models (technically outside our scope), such as the cubic Dresselhaus interaction, do not give rise to a spin polarization in an electric field, most models describing 2D systems in diamond and zincblende lattices do. We therefore expect in-plane spin-polarized phases with a shifted Fermi surface to occur generally in 2D systems with strong spin-orbit interactions. We note, however, that in addition to the requirement of gyrotropic symmetry, it is also necessary for the system to have two Fermi surfaces in the non-interacting state. Systems such as topological insulators, in which the spin-orbit interaction is dominant and have a single Fermi surface, are not expected to exhibit in-plane spin-polarized phases. 

\subsection{Experimental detection}

We would like to discuss the possibility of observing the in-plane spin-polarized phase, the most unconventional phase predicted by our work, in the laboratory. Given the shift in the Fermi surface and the existence of a net spin-orbit effective field it also follows that in the in-plane spin polarized phase a spatial direction is preferred and rotational symmetry is broken. When a small in-plane external magnetic field is applied we expect an anisotropy in the charge current as the magnetic field is rotated in the plane of the 2DEG. The Fermi surface shift likewise introduces a new characteristic wave vector in the system and this could in principle be detected by point-contact interferometry \cite{Katine_PRL97}. 

Noting that the in-plane phase displays an unconventional magnetization, we recall that one of the most reliable probes of a magnetized system is the occurrence of the anomalous Hall effect, which does not require an external magnetic field. The experimental setup to detect this effect is straightforward. However, due to the shift in the Fermi surface and the complex in-plane spin texture the calculation of the anomalous Hall conductivity will need to be performed as a separate project. 

Finally, experimental realization of the state we discuss would create magnetic structures in the absence of any doping with magnetic impurities, utilizing instead the electric-field-tunable Rashba spin-orbit coupling. Such systems could become building blocks for novel spintronic devices and platforms for realizing Majorana fermions.

\section{Summary of results}

In the first part of our analysis we demonstrated that, when an interacting Rashba spin-orbit coupled system is placed in an external electric field, the current-induced spin polarization diverges at a certain interaction strength, while the charge current is unaffected by electron-electron interactions. Based on this insight we concluded that an in-plane spin polarized phase can exist in equilibrium in this system, in which the Fermi surface is shifted away from the zone centre and as a consequence a net spin-orbit effective field exists. 

In the second part we established the complete mean-field phase diagram of a Rashba spin-orbit coupled system in the presence of electron-electron interactions. We recovered an out-of-plane spin polarized phase found previously, as well as the expected in-plane spin-polarized phase. The Fermi surface of the system is shifted from the centre of the Brillouin zone, and it displays a variety of spin textures, which depend on the strength of the spin-orbit interaction. The in-plane spin polarized phase we have identified is akin to the Pomeranchuk instability. At low interaction strengths we found two expected Fermi liquid phases, one with a single Fermi surface and one with two Fermi surfaces, and mapped out the Lifshitz transition between them.

\section{Outlook}

Establishing the mean-field phase diagram is the customary first step when approaching strongly-correlated problems. Bearing in mind that the Hartree-Fock approximation tends to overestimate the exchange energy and underestimate $r_{s}$ for the Bloch transition \cite{PhysRevLett.45.566}, the natural extension of the theory involves going beyond mean-field to test our results qualitatively and quantitatively by (i) performing a random-phase approximation calculation and (ii) devising a reliable method to include the correlation energy. Typically, the inclusion of screening tends to shift the phase boundaries to larger values of $r_s$, but we do not expect the topology of the phase diagram to change. We note that large values of $r_s$ have been reported in semiconductor nanostructures \cite{Simmons_PRL98, Hanein_PRB98}.  

The fact that the driving force behind the in-plane spin polarized phase is the spin-orbit interaction by itself gives rise to two important questions. Firstly, it is important to determine what forms of spin currents, if any, are associated with the in- and out-of plane phases. It is well known that spin currents, at least when using the conventional definition, can exist in thermodynamic equilibrium \cite{PhysRevB.68.241315}, and the possibility exists that spin eddy currents could circulate in the spin polarized phases. Secondly, it has long been known that the form of the spin-orbit interaction can be tailored by the material growth direction \cite{Winkler03}. An interesting open problem concerns the possible spin-polarized phases associated with unconventional forms of the spin-orbit coupling that lack the symmetries of the Rashba model, such as its rotational symmetry. 

In this context, in a future publication we will study the interplay of electron-electron interactions and spin-orbit coupling in a 2D electron gas in a semiconductor with both linear Rashba and linear Dresselhaus spin-orbit interactions. When both Rashba and Dresselhaus are present and are of equal magnitude the effective magnetic field describing the spin-orbit interaction singles out a well-defined direction in momentum space \cite{Bernevig06prl}. The noninteracting ground state already has shifted Fermi surfaces and the spins point in a well-defined direction, hence interactions are expected to stabilize a state with an in-plane spin polarization. The limit in which the Rashba and Dresselhaus interactions are equal in magnitude has been of interest because of the fixed direction of the momentum-dependent spin-orbit magnetic field and the occurrence of the persistent spin helix, which has been realized experimentally \cite{Koralek_SpinHelix_Nature09}.

We expect likewise a rich phase diagram in 2D spin-3/2 hole systems, which exhibit very strong spin-orbit coupling having a nontrivial functional form, a complex sub-band structure with several anticrossings, and a large Wigner-Seitz radius $r_s$ even at relatively high densities. Based on the findings of this work many possibilities exist for magnetic ground states that may be observed experimentally. 

\acknowledgements

We would like to acknowledge M.~A.~Eriksson for useful discussions. This research was supported by the Australian Research Council Centre of Excellence in Future Low-Energy Electronics Technologies (project number CE170100039) and funded by the Australian Government. SC acknowledges support from the National Key Research and Development Program of China (Grant No. 2016YFA0301200) and the NSFC grants (No. 11574025 and No. U1530401). He is also indebted to the late Gabriele F. Giuliani, under whose supervision some of the results presented here were obtained \cite{Chesi_PhD}. RW was supported by the NSF under grant No.\ DMR-1310199.

\bibliographystyle{apsrev4-1} 
% \bibliography{Rashba_ee}


\end{document}